\documentclass[10pt,aps,pra,
               twocolumn,
               nofootinbib,
               floatfix,
			   superscriptaddress,
               showpacs
              ]{revtex4-1}

\usepackage{graphicx}
\usepackage{dcolumn}
\usepackage{bm}
\usepackage[utf8]{inputenc}
\usepackage{amssymb}
\usepackage{amsmath}
\usepackage{pstricks}
\usepackage{color}
\usepackage{slashed}
\usepackage{braket}
\usepackage{hyperref}
\usepackage{natbib}
\usepackage{extarrows}
\usepackage{titlesec}
\usepackage{wasysym}
\usepackage[normalem]{ulem}

\newcommand{\figref}[1]{Fig.~\ref{#1}}
\newcommand{\EDfigref}[1]{Fig.~\ref{#1}}
\newcommand{\Eqref}[1]{Eq.~\eqref{#1}}

\newcommand{\redbullet}{\protect\includegraphics[height= 6 pt]{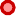}}
\newcommand{\bluediamond}{\protect\includegraphics[height= 8 pt]{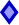}}
\newcommand{\purplesquare}{\protect\includegraphics[height= 6 pt]{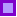}}

\definecolor{ao(english)}{rgb}{0.0, 0.5, 0.0}
\definecolor{applegreen}{rgb}{0.55, 0.71, 0.0}
\definecolor{cadetblue}{rgb}{0.37, 0.62, 0.63}
\definecolor{cadet}{rgb}{0.33, 0.41, 0.47}
\definecolor{byzantine}{rgb}{0.74, 0.2, 0.64}
\definecolor{darkgreen}{rgb}{0.0,0.4,0.0}
\definecolor{plottingpurple}{rgb}{0.5,0,0.8}
\definecolor{plottingred}{rgb}{0.8,0,0}
\definecolor{plottinggreen}{rgb}{0,0.7,0.2}
\definecolor{plottingblue}{rgb}{0,0,0.8}

\renewcommand{\i}{\mathrm{i}}
\newcommand{\e}{\mathrm{e}}

\newcommand{\eq}[1]{(\ref{eq:#1})}
\newcommand{\Eq}[1]{Eq.~(\ref{eq:#1})}

\newcommand{\Sect}[1]{Sect.~\ref{sec:#1}}

\definecolor{ao(english)}{rgb}{0.0, 0.5, 0.0}

\newcolumntype{L}[1]{>{\raggedright\let\newline\\\arraybackslash\hspace{0pt}}m{#1}}
\newcolumntype{C}[1]{>{\centering\let\newline\\\arraybackslash\hspace{0pt}}m{#1}}
\newcolumntype{R}[1]{>{\raggedleft\let\newline\\\arraybackslash\hspace{0pt}}m{#1}}

\makeatletter
\let\cat@comma@active\@empty
\makeatother

\begin{document}

\title{{On solving quantum many-body problems by experiment}}

\author{Thomas Schweigler} 
\affiliation{Vienna Center for Quantum Science and Technology, Atominstitut, TU Wien, Stadionallee 2, 1020 Vienna, Austria}

\author{Valentin Kasper}
\affiliation{Institut f\"{u}r Theoretische Physik, Universit\"{a}t Heidelberg, Philosophenweg 16, 69120 Heidelberg, Germany}

\author{Sebastian Erne}
\affiliation{Vienna Center for Quantum Science and Technology, Atominstitut, TU Wien, Stadionallee 2, 1020 Vienna, Austria}
\affiliation{Institut f\"{u}r Theoretische Physik, Universit\"{a}t Heidelberg, Philosophenweg 16, 69120 Heidelberg, Germany}

\author{Igor Mazets}
\affiliation{Vienna Center for Quantum Science and Technology, Atominstitut, TU Wien, Stadionallee 2, 1020 Vienna, Austria}
\affiliation{Wolfgang Pauli Institute, 1090 Vienna, Austria}

\author{Bernhard Rauer}
\affiliation{Vienna Center for Quantum Science and Technology, Atominstitut, TU Wien, Stadionallee 2, 1020 Vienna, Austria}

\author{Federica Cataldini}
\affiliation{Vienna Center for Quantum Science and Technology, Atominstitut, TU Wien, Stadionallee 2, 1020 Vienna, Austria}

\author{Tim Langen}
\altaffiliation[Present address: ]{5. Physikalisches Institut \& Center for Integrated Quantum Science and Technology, Universit\"at Stuttgart, 70569 Stuttgart, Germany}  
\affiliation{Vienna Center for Quantum Science and Technology, Atominstitut, TU Wien, Stadionallee 2, 1020 Vienna, Austria}
 
\author{Thomas Gasenzer}
\affiliation{Kirchhoff-Institut f\"{u}r Physik, Universit\"{a}t Heidelberg, Im Neuenheimer Feld 227, 69120 Heidelberg, Germany}

\author{J\"urgen Berges}
\affiliation{Institut f\"{u}r Theoretische Physik, Universit\"{a}t Heidelberg, Philosophenweg 16, 69120 Heidelberg, Germany}
        
\author{J\"org Schmiedmayer}
\email[]{schmiedmayer@atomchip.org}
\affiliation{Vienna Center for Quantum Science and Technology, Atominstitut, TU Wien, Stadionallee 2, 1020 Vienna, Austria}

\date{14 December 2016}


\maketitle

{\bf
Knowledge of all correlation functions of a system is equivalent to solving the corresponding many-body problem~\cite{Schwinger1951A,Schwinger1951B}. Already a finite set of correlation functions can be sufficient to describe a quantum many-body system if correlations factorise, at least approximately. While being a powerful theoretical concept, an implementation based on experimental data has so far remained elusive. 
Here, this is achieved by applying it to a non-trivial quantum many-body problem: A pair of tunnel-coupled one-dimensional atomic superfluids.
From measured interference patterns we extract phase correlation functions up to $\mathbf{10^{\mathrm{th}}}$ order and analyse if, and under which conditions, they factorise. This characterises the essential features of the system, the relevant quasiparticles, their interactions and possible topologically distinct vacua. We verify that in thermal equilibrium the physics can be described by the quantum sine-Gordon model~\cite{Coleman75,Mandelstam,Faddeev19781,Sklyanin1979}, relevant for a wide variety of disciplines from particle to condensed-matter physics~\cite{cuevas2014sine,fogel1977dynamics,Gritsev07}. Our experiment establishes a general method to analyse quantum many-body systems in experiments. It represents a crucial ingredient towards the implementation and verification of quantum simulators~\cite{Cirac2012}.
}
 
One central objective of quantum field theories is to capture the essential physics of complex quantum many-body systems in terms of collective degrees of freedom~\cite{bogoliubov1947theory, landau1957theory}. Their propagation and interaction are encoded in the  
correlation functions
\begin{equation}
	G^{(N)}({\bf{z}}) = \braket{ \mathcal{O}(z_1) \mathcal{O}(z_2) \dots \mathcal{O}(z_N) } ,
	\label{eq:CorrFunc}
\end{equation}
where $\mathcal{O}(z_i)$ are the corresponding Heisenberg operators 
evaluated at coordinate $z_i$, and $N$ is the order of the correlation. 
$G^{(N)}$ can be decomposed into~\cite{gardiner2009stochastic,zinn2002quantum} 
\begin{equation}
	G^{(N)}({\bf{z}}) = G_{\mathrm{dis}}^{(N)}({\bf{z}}) + G_{\mathrm{con}}^{(N)}({\bf{z}}) ~\mathrm{.}
	\label{eq:DecomposeCorrFunc}
\end{equation}
The first term $G_{\mathrm{dis}}^{(N)}$ is the {\em disconnected} part of the correlation function. It is fully determined by the lower-order correlation functions and therefore does not contain new information at order $N$.  

The second term, $G_{\mathrm{con}}^{(N)}$, is the {\em connected} part of the correlation function, and contains genuine new information about the system at order $N$. In a diagrammatic expansion, $G^{(N)}_{\mathrm{con}}$ is given by a sum of fully connected diagrams with $N$ external lines.

If $G_{\mathrm{con}}^{(N)}$ is zero for all $N > 2$, the higher-order correlation functions factorise and are given by the Wick decomposition~\cite{Wick1950,zinn2002quantum} containing only terms of $G^{(N)}$ with $N \le 2$. In this case, the quantum many-body states are Gaussian. Determining the collective degrees of freedom that lead to complete factorisation corresponds to solving the quantum many-body problem. 

Finding out up to which order $N$ the connected correlation function can be estimated with statistical significance gives a direct handle on how much of the complexity of the underlying quantum many-body system is accessible in a given experiment. 
	
\begin{figure}[b!]
	\centering
	\includegraphics[width=\columnwidth]{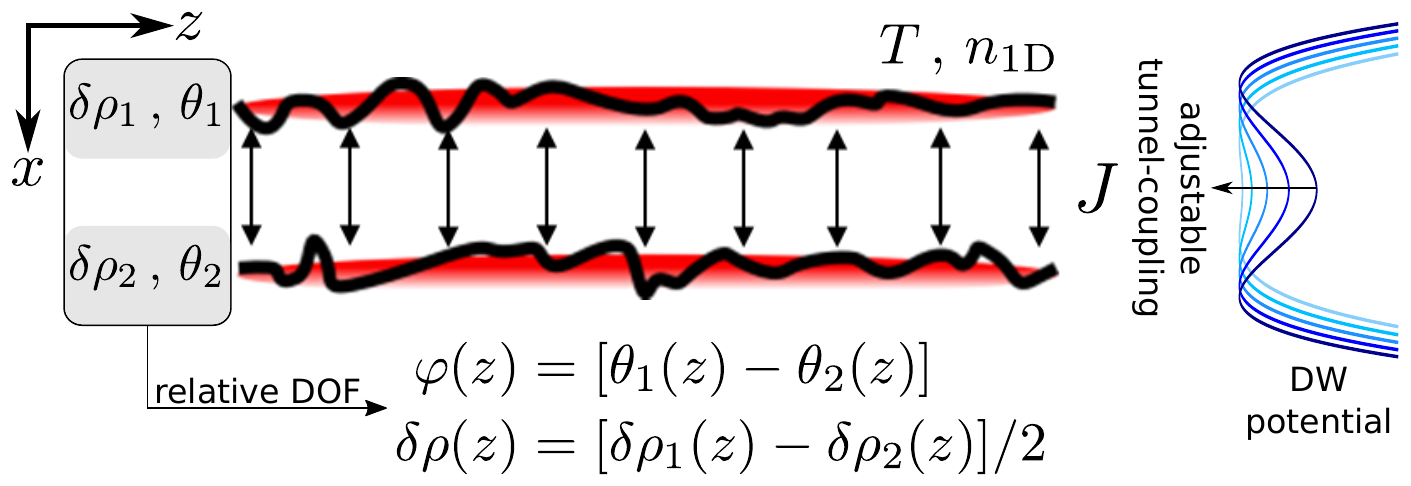}
	\caption{{\bf Schematics of the experimental setup.} We consider two tunnel-coupled one-dimensional superfluids in a double-well (DW) potential at a common temperature $T$. Changing the barrier height of the DW potential (blue lines) allows for an adjustable tunnel-coupling $J$ between the two superfluids. The superfluids are described in terms of density fluctuations $\delta\rho_{1,2}$ around their equal mean densities $n_{\mathrm{1D}}$ and fluctuating phases $\theta_{1,2}$ (black lines). From these quantities we define the relative degrees of freedom (DOF) $\delta\rho$ and $\varphi$ used for the discussion in the main text.
	}
	\label{fig:geometry}
\end{figure}

With the rapid progress in quantum gas experiments ~\cite{Bloch2012}, measuring higher-order correlation functions~\cite{Hodgman2011,Dall2013,Endres2013,Langen15} is now within reach. To illustrate the power of the above concepts to analyse a non trivial interacting quantum many-body system, we experimentally investigate two tunnel-coupled one-dimensional (1D) bosonic superfluids, realised with quantum degenerate $^{87}\mathrm{Rb}$ atoms trapped in a double-well (DW) potential with a freely adjustable barrier (\figref{fig:geometry}) \cite{Betz2011}. Matter-wave interferometry~\cite{Schumm05,Gring12,Langen13b} gives direct access to the spatially resolved relative phase $\varphi (z)$ between the superfluids (see Methods). 

Tunneling through the DW-barrier drives the relative phase $\varphi(z)$ towards zero. The strength of this `phase locking' is characterised by $\braket{\cos(\varphi)}$, a quantity that is zero for completely random phases (no phase locking) and approaches unity in the limit of strong phase locking. The value of $\braket{\cos(\varphi)}$ depends on the tunnel-coupling strength as well as on the temperature \cite{Betz2011}. 	

From the measured phase profiles $\varphi (z)$ we extract the $N^{\mathrm{th}}$-order correlation functions of the phase by evaluating  
\begin{equation}
G^{(N)}({\bf{z}},{\bf{z}}') = \braket{[\varphi(z_1)- \varphi(z'_1)]\dots[\varphi(z_N) - \varphi(z'_N)]} ,
\label{eq:CorrelationFunction}
\end{equation}
with coordinates ${\bf{z}}=(z_1, \dots, z_N)$ and ${\bf{z}}'=(z'_1,\dots,z'_N)$ along the length of the system. The brackets $\braket{\ldots}$ denote averaging over many experimental realisations. For details on how to calculate $G_{\mathrm{con}}^{(N)}$ and $G_{\mathrm{dis}}^{(N)}$ see the Methods section.

\begin{figure}[tb!]
	\centering
	\includegraphics[width=\columnwidth]{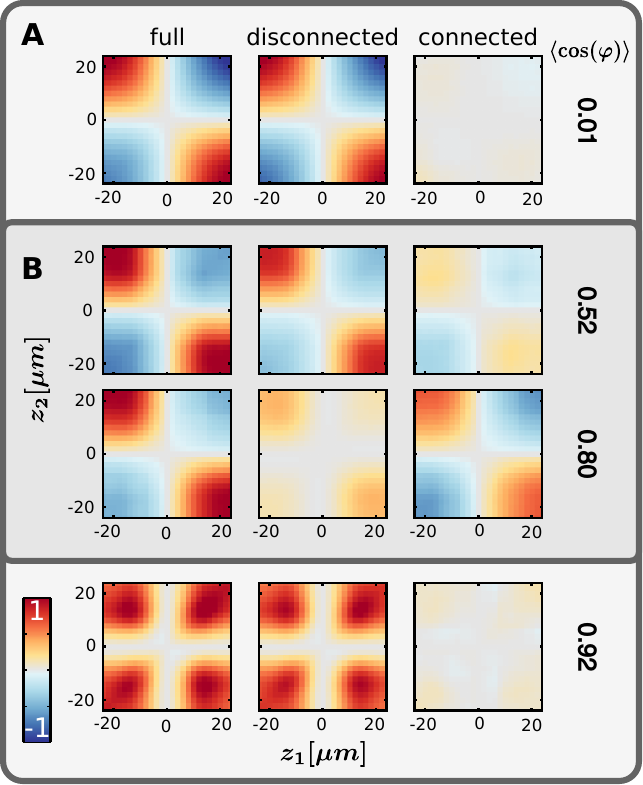}
	\caption{{\bf Decomposition of the \boldmath{$4^{\mathrm{th}}$}-order phase correlation functions $G^{(4)}({\bf{z}},{\bf{z}}')$.}  ({\bf A}) uncoupled ($\braket{\cos(\varphi)} \simeq 0$) and strongly phase-locked ($\braket{\cos(\varphi)} \simeq 1$); ({\bf B}) intermediate regime. To visualise the high-dimensional data, we choose $z_3 = -z_4 = 14\ \mu$m and ${\bf{z}}' = 0$, leading to the observed symmetric crosses where the correlation function vanishes. The colour marks the value of the full, disconnected and connected correlation functions, with each row normalised to its maximum value, implying colour to encode the interval from $-1$ to $1$. }
	\label{fig:wick_visual}
\end{figure}

\figref{fig:wick_visual} shows the experimental data for the full $4^{\mathrm{th}}$-order correlation function $G^{(4)}({\bf{z}},{\bf{z}}')$, its disconnected, and connected parts, for different strengths of the phase locking between the superfluids. The superfluids are prepared by slow evaporative cooling into the DW potential with the aim of creating a thermal equilibrium state. In both limits, $\braket{\cos(\varphi)} \simeq 0$ (uncoupled superfluids) and $\braket{\cos(\varphi)} \simeq 1$ (strongly coupled superfluids), the connected part vanishes (\figref{fig:wick_visual}A). The full $4^{\mathrm{th}}$-order correlation function is given by its disconnected part, calculated from the $2^{\mathrm{nd}}$-order correlation function, i.e., the $4^{\mathrm{th}}$-order correlation function factorises. For intermediate phase locking (\figref{fig:wick_visual}B) the $4^{\mathrm{th}}$-order function cannot be described by $2^{\mathrm{nd}}$-order functions alone, and a significant connected part remains.

We now compare our observations with predictions for thermal states of the sine-Gordon model, which has been proposed as an effective description for the relative degrees of freedom of two tunnel-coupled one-dimensional (1D) bosonic superfluids~\cite{Gritsev07}. Following~\cite{Gritsev07} (for details see the SI) the Hamiltonian is given by 
\begin{align} 
\begin{split}
\label{eq:SG}
{H}_{\mathrm{SG}} =   &\int \mathrm{d}z \left[ g \delta {{\rho}}^2 + \frac{\hbar^2 n_\mathrm{1D}}{4m} (\partial_z {{\varphi}})^2 \right] \\
&- \int{ \mathrm{d}z ~ 2 \hbar J n_\mathrm{1D} \cos({\varphi}) } \, \mathrm{,}
\end{split}
\end{align}
where $\delta \rho(z)$ are the relative density fluctuations and $\varphi(z)$ is the relative phase (see \figref{fig:geometry}). These fields represent canonically conjugate variables fulfilling appropriate commutation relations. The parameter $m$ is the mass of the atoms, $g$ is the 1D interaction strength, and $J$ the tunnel-coupling strength between the superfluids with equal 1D-densities $n_{\mathrm{1D}}$.

The correlation functions in \Eqref{eq:CorrelationFunction} reflect the correlations in the collective degrees of freedom, constructed from the conjugate fields $ \delta \rho $ and $\varphi$. The connected correlation function $G_{\mathrm{con}}^{(N)}({\bf{z}},{\bf{z}}')$ for $N>2$ is therefore a direct measure of their interactions. In contrast, the more commonly used correlation functions $\braket{e^{i[\varphi(z)-\varphi(z')]}}$ and their higher-order generalisations \cite{Langen15} contain $G_{\mathrm{con}}^{(N)}({\bf{z}},{\bf{z}}')$ up to arbitrary order even for the $2^{\mathrm{nd}}$-order function and are therefore not suitable for studying these interaction properties (see Methods and the SI).

${H}_{\mathrm{SG}}$ nicely reflects the observations in \figref{fig:wick_visual}. For $\braket{\cos(\varphi)} \simeq 0$, corresponding to $J \simeq 0$, only the first part of $H_{\mathrm{SG}}$, the quadratic Tomonaga-Luttinger Hamiltonian~\cite{Tomonaga1950a, Luttinger1963a, Lieb1965}, remains, leading to Gaussian thermal states characterised by a vanishing connected correlation function $G^{(N)}_{\mathrm{con}}$ for $N>2$. For $\braket{\cos(\varphi)} \simeq 1$ we can replace the cosine in the Hamiltonian \eqref{eq:SG} by its harmonic approximation~\cite{whitlock2003relative} leading as well to a quadratic Hamiltonian and Gaussian fluctuations. 

For intermediate phase locking (intermediate $\braket{\cos(\varphi)}$) we have to consider the full cosine potential leading to a non-vanishing $4^{\mathrm{th}}$-order connected correlation function.

We emphasise that the Hamiltonian \eqref{eq:SG} represents an effective low-energy description for the underlying microscopic degrees of freedom and processes. Theoretically, the insensitivity to details of the underlying micro-physics can be efficiently phrased in terms of relevant and irrelevant operators of the model employed. The observed factorisation for strong and vanishing tunnel-coupling provides a clean experimental demonstration that the contributions from a vast set of possible irrelevant operators renormalise to zero in the low-energy effective theory describing thermal equilibrium.

\begin{figure}[tb!]
	\includegraphics[width=\columnwidth]{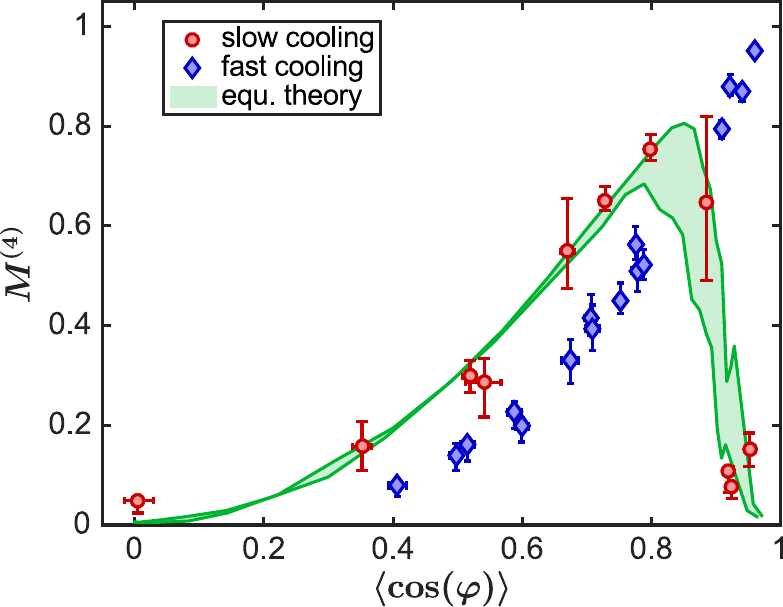}
	\caption{ {\bf Relative size of the \boldmath{$4^{\mathrm{th}}$}-order connected correlation function,} for thermal equilibrium prepared by slow evaporative cooling (\redbullet) and for a non-thermal state reached by fast cooling (\bluediamond). Plotted is the measure $M^{(4)}$ (defined in \Eqref{eq:int_measure}) vs.\ the phase locking strength quantified by $\braket{\cos(\varphi)}$. One sees good agreement between thermal equilibrium data and thermal sine-Gordon theory. The fast-cooling data clearly deviates from the equilibrium theory prediction. The theory curves (\textcolor{plottinggreen}{green shaded region}) were calculated for the maximum spread of the estimated experimental parameters. The error bars represent 80\% confidence intervals calculated by using bootstrapping (see Methods).
	}
	\label{fig:integrated_thermal}
\end{figure}

For a quantitative comparison between experiment and equilibrium sine-Gordon theory we first estimate the density $n_\mathrm{1D}$ and the temperature $T$ of our samples from independent measurements (see Methods). We then calculate numerically the theoretical prediction for the higher-order correlation functions (see Methods and the SI). We compare theory and experiment using the measure
\begin{equation}
M^{(N)}=\frac{\sum_{\bf{z}}{\left|G^{(N)}_{\mathrm{con}}({\bf{z}},0)\right|}}{\sum_{\bf{z}}{\left|G^{(N)}({\bf{z}},0)\right|}} \label{eq:int_measure}
\end{equation}
plotted in \figref{fig:integrated_thermal} as a function of the phase locking strength quantified by $\braket{\cos(\varphi)}$. The experimental results for $N = 4$ agree well with the sine-Gordon equilibrium theory.

Looking at the Wick decomposition for the $6^{\mathrm{th}}$-, $8^{\mathrm{th}}$- and $10^{\mathrm{th}}$-order functions, i.e.\ whether they factorise into $2^{\mathrm{nd}}$-order functions, one gets similar results (see \EDfigref{fig:edfigwick6p8p}). For $\braket{\cos(\varphi)} \simeq 0$ they can be described by $2^{\mathrm{nd}}$-order functions only; in the intermediate regime this is not possible; towards  $\braket{\cos(\varphi)} \simeq 1$ factorisation can be achieved, but conditions get more stringent with increasing order N. 

Experimentally measuring the $6^{\mathrm{th}}$-, $8^{\mathrm{th}}$- and $10^{\mathrm{th}}$-order (or higher) \emph{connected} correlation functions, i.e.\ investigating their factorisation into all lower-order correlation functions, is a much more challenging task. Factorisation for very weak and very strong phase locking follows from the observed validity of the Wick decomposition for these cases. In the intermediate regime, the relative size of the connected part is statistically significant and there is qualitative agreement between experiment and thermal equilibrium theory (see Figs.~\ref{fig:edfig6pm}--\ref{fig:edfig10pslices}). These non-vanishing connected parts are a clear indication that in our system three-, four- and five-particle interactions are important. This highlights how our method could provide a new access to effective few-body microscopic aspects of many-body dynamics.

We want to emphasize that, to arrive at the sine-Gordon model from the original Hamiltonian describing two tunnel-coupled 1D superfluids, one has to go through a series of approximations that lead to the decoupling of the symmetric and anti-symmetric modes of the system (see the SI). These approximations include only terms second-order in $\delta \rho$ and $\partial_z \varphi$, and neglect mixed density-phase terms. Showing that the measured correlation functions up to $10^\mathrm{th}$ order (containing terms up to $\varphi^{10}$)  are faithfully reproduced by $H_{SG}$ demonstrates that the approximations needed to derive this low-energy effective theory are justified, at least in equilibrium.

So far we discussed data for the system prepared by very slow cooling which can be described by the thermal equilibrium sine-Gordon theory. Systems prepared by a final cooling speed a factor of 10 faster (see Methods) show a different behaviour (\figref{fig:integrated_thermal}). This demonstrates that our method can differentiate between thermal and non-thermal states.  

It is interesting to point out that for strong phase locking ($\braket{\cos(\varphi)} \simeq 1$) a significant connected part remains in the fast cooled sample, indicating that in the non-thermal case the cosine in the Hamiltonian \eqref{eq:SG} is relevant even in this regime.

\begin{figure}[tb!]
	\includegraphics[width=\columnwidth]{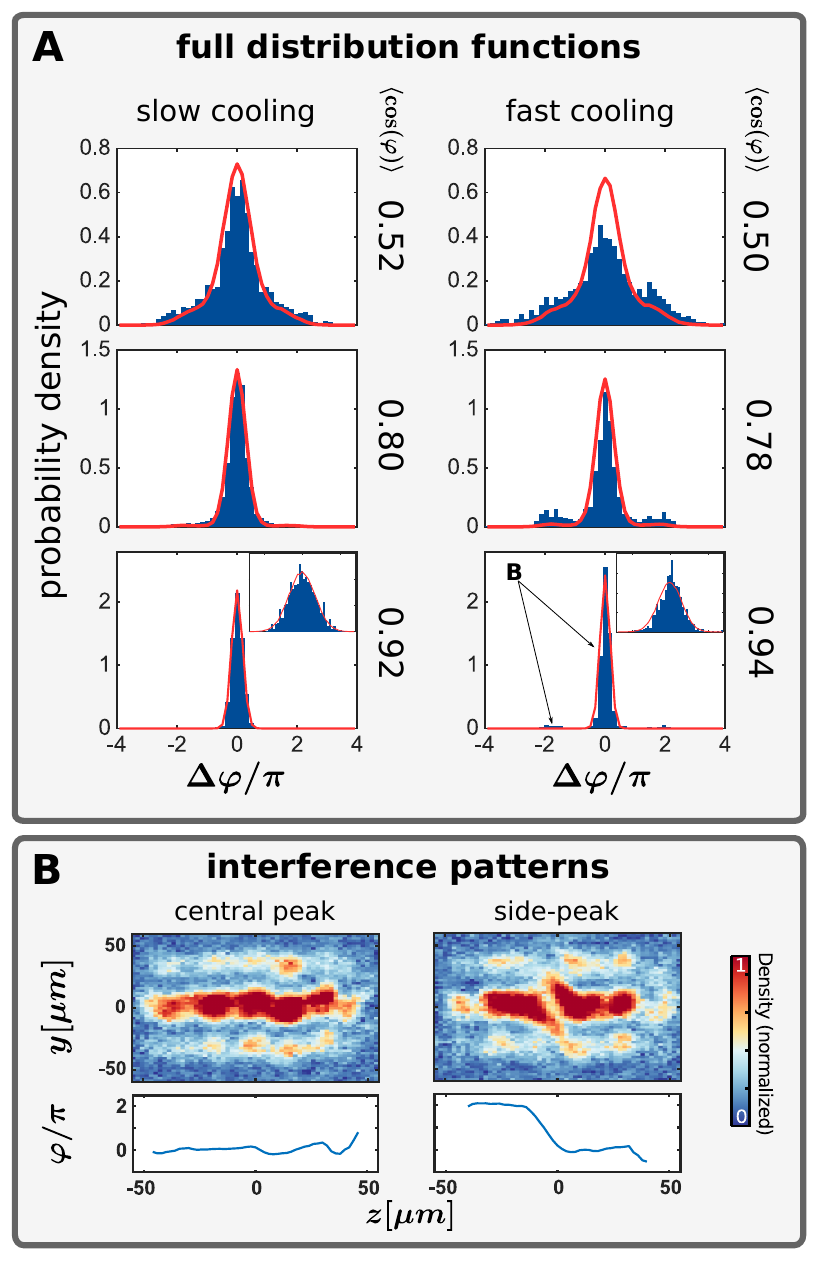}
	\caption{ {\bf Full distribution functions and interference patterns of the phase.} \textbf{(A)} Full distribution functions for the phase differences $\Delta\varphi = \varphi(z)-\varphi(z')$ for $z = -z' = 20\ \mu \mathrm{m}$ for different phase-locking strengths and two different ways to prepare the quantum gas: (\emph{left}) very slow cooling (\emph{right}) fast cooling (see main text for details). The experimental data (blue bars) for the system prepared by slow cooling are in good agreement with the thermal sine-Gordon theory (red lines). The rapidly cooled systems show significant deviations, with especially pronounced peaks at $\pm 2 \pi$. \textbf{(B)}~Interference patterns of the phase for the strongly coupled fast cooled system contributing to the central and side peaks of the full distribution function. In the central peak sample, phase fluctuations are small, whereas, in the side peak sample,  a sine-Gordon soliton is clearly visible.
		 }
	\label{fig:PhaseDist}
\end{figure}

To gain insight into the mechanisms leading to the difference between slow and fast cooling, we analyse the full distribution function of the phase differences $\Delta\varphi = \varphi(z)-\varphi(z')$ to which, in principle, all $N^{\mathrm{th}}$-order phase correlation functions contribute. \figref{fig:PhaseDist}A shows the full distributions for one particular pair of coordinates $(z, z')$ chosen symmetrically around the center of the trap. 

For \emph{slow cooling} and intermediate values of $\braket{\operatorname{cos}(\varphi)}$ the full distribution functions of the phase differences $\Delta\varphi$ are distinctly non-Gaussian. For strong phase locking ($\braket{\operatorname{cos}(\varphi)} \simeq 1$) we find Gaussian full distribution functions, as anticipated from the observed validity of the Wick decomposition in this case. In contrast, for \emph{fast cooling} all coupled cases show non-Gaussian distribution functions. With increasing phase locking, one can see the appearance of distinct side peaks at $\pm 2\pi$, becoming more localised, but at the same time more suppressed. 

For $\braket{\operatorname{cos}(\varphi)} = 0.94$ we observe a Gaussian central peak (see inset) as well as a few outliers at $\pm 2\pi$. Studying interference patterns for individual realisations corresponding to the side peaks reveals that the phase rotates through a full circle of $2\pi$ within a short distance (see \figref{fig:PhaseDist}B). These localised kinks represent transitions between different minima of the cosine potential and can be identified as solitons of the sine-Gordon model; they are topological excitations of $H_{\mathrm{SG}}$ (\Eqref{eq:SG}). 

In the case of fast cooling these sine-Gordon solitons are frozen in, and the phase of the quantum field fluctuates around them. Such states may therefore be interpreted as topologically distinct, `false' vacua~\cite{Coleman77A} above which the quasiparticles are being excited. The energy of these false vacua increases with the number of sine-Gordon solitons.

The procedure outlined in our letter is the basis for a general principle to extract information from non-trivial quantum systems in an unbiased and unambiguous way. It represents an important step towards solving complex quantum many-body problems by experiment. Higher-order correlation functions hold furthermore large promise for experimental and theoretical investigations of non-equilibrium dynamics. 
Our method thus provides a new and important tool for future quantum simulators.


\vspace{4mm}
\noindent{\bf Acknowledgements:} \\
We acknowledge discussions with E.\ Demler, and J.~M.\ Pawlowski. This work was supported by the EU through the EU-FET Proactive grant AQuS, Project No. 640800, and the ERC advanced grant QuantumRelax, and by the Austrian Science Fund (FWF) through the doctoral programme CoQuS (\textit{W1210}) (T.S., B.R., F.C.), and the SFB-FoQuS. This work is part and supported by the SFB 1225 `ISOQUANT' financed by the German Research Foundation (DFG) and the FWF. V.K. acknowledges support from the Max Planck Society through the doctoral programme IMPRS-QD. T.L. acknowledges support by the Alexander von Humboldt Foundation through a Feodor Lynen Research Fellowship. T.G. and J.B. acknowledge support by the University of Heidelberg (Center for Quantum Dynamics) and the Helmholtz Association (HA216/EMMI).

\vspace{4mm}
\noindent{\bf Author contributions:} \\
T.S., B.R., F.C.\ and T.L.\ performed the experiment and the data analysis. S.E., V.K., T.S.\ and I.M.\ did the theoretical calculations. J.S., J.B.\ and T.G.\ provided scientific guidance in experimental and theoretical questions. All authors contributed to the interpretation of the data and the writing of the manuscript.

\bibliographystyle{mynaturemag}
\bibliography{bibliofinal}

\vspace{4mm}




\renewcommand{\thesection}{}
\titlelabel{}
\section{Methods}
\titlelabel{\thetitle. }


\subsection{Preparation of the coupled 1D superfluids}
The coupled one-dimensional (1D) superfluids are realised using our standard procedure to produce ultracold gases of ${}^{87}$Rb in a double-well  potential on an atom chip. Each well consists of a highly elongated cigar-shaped trap (two tightly confined directions, one direction with weak confinement). The wells are separated along one of the tightly confined directions (see \figref{fig:geometry}). The separation is horizontal, avoiding the influence of gravity. By tuning the height of the barrier separating the wells we can change the tunnel-coupling between the two superfluids. 

The clouds are prepared by evaporatively cooling the atoms whilst keeping the double-well trap static. The relative evaporation rates at the end of the cooling ramp amount to a few percent per 10~ms for slow cooling and a few 10\% for fast cooling. In both cases, the measurements have been performed right after the evaporative cooling.

The single wells have measured harmonic frequencies of $\omega_\perp \simeq 2\pi\times1.4\,$kHz in the radial direction and $\omega_z \simeq 2\pi\times6.7\,$Hz in the longitudinal direction. The temperature, atom number and chemical potential are $T=11 \ldots56\,$nK, $N=4000 \ldots 6300$ per well, and $\mu=2 \pi \hbar\times (0.70 \ldots0.94)\,$kHz, respectively, such that the 1D condition $\mu, k_B T < \hbar \omega_\perp$ is well fulfilled within each well. The gas typically has a length of about $100\,\mu$m, from which we use the central $50\, \mu$m (density variation of about $25 \%$) for our analysis.

\subsection{Measurement of the relative phase}
To extract the spatially resolved relative phase between the superfluids we record the resulting matter-wave interference pattern of the two 1D Bose gases after about $16\,$ms time-of-flight expansion using standard absorption imaging~\cite{Gring12,Smith13}. 

The local position of the fringes in the fluctuating interference pattern directly corresponds to the relative in-situ phase $\varphi(z)$ between the two superfluids. We extract this relative phase by fitting a sinusoidal function to each pixel line in the interference pattern~\cite{Langen13b}. This allows to determine $\varphi(z)$ modulo $2 \pi$. However, assuming that the phase difference between two neighboring points along the $z$-direction is within the interval $(-\pi,\pi]$, one can reconstruct the full phase profiles to obtain unambiguous phase differences $\varphi(z) - \varphi(z')$ not restricted to $(-\pi,\pi]$. The remaining global ambiguity of $2 \pi n$ ($n$ is an integer number) for the phase field $\varphi(z)$ is irrelevant for our analysis. A sample of $\varphi(z)$ is generated by repeating the experiment $290 \ldots 2800$ times (typically 1000 times).

\subsection{Calculation of the correlation functions and their connected parts}
As defined in \Eq{CorrelationFunction} in the main text, we evaluate the phase correlation functions as
\begin{equation*}
G^{(N)}({\bf{z}},{\bf{z}}') = \braket{[\varphi(z_1)- \varphi(z'_1)]\dots[\varphi(z_N) - \varphi(z'_N)]},
\end{equation*}
where the brackets denote the averaging over different experimental realisations and $\varphi(z)$ is extracted as described in the previous subsection. Similarly, we calculate the connected part using~\cite{shiryaev2016probability}
\begin{align}
\begin{split}
G_{\mathrm{con}}^{(N)}({\bf{z}},{\bf{z}}') =  \sum_{\pi} \Bigg[ \ &(|\pi|-1)! \ (-1)^{|\pi|-1}  \\
 &\prod_{B \in \pi} \left\langle \prod_{i \in B} [\varphi(z_i)- \varphi(z'_i)] \right\rangle \Bigg]  . \label{general_formula_connected}
\end{split}
\end{align}
Here the sum runs over all possible partitions $\pi$ of $\{1,\dots,n\}$, the first (left) product runs over all blocks $B$ of the partition and the second (right) product runs over all elements $i$ of the block. $|\pi|$ is the number of blocks in the partition. The number of partitions is given by the Bell number $B_N$, which quickly grows with $N$. For $N=10$ (the highest order investigated in this paper) we get $B_{10} = 115975$, for the next even order it would already be $B_{12} = 4213597$. Using the central moments in \Eqref{general_formula_connected}, all partitions containing blocks of size one do not contribute, this significantly reduces the number of terms in the sum. However, the computational effort still rises quickly with the order $N$. Note that \Eqref{general_formula_connected} does not represent an unbiased estimator of the connected correlation function. However, for our large sample sizes the bias should be negligible.

Let us also explicitly write down the Wick decomposition 
\begin{align}
\begin{split}
	G_{\mathrm{wick}}^{(N)}({\bf{z}},{\bf{z}}') =  \sum_{\pi_2} \Bigg[ \prod_{B \in \pi_2} \Big\langle &[\varphi(z_{B_1})- \varphi(z'_{B_1})] \\
	 &[\varphi(z_{B_2})- \varphi(z'_{B_2})] \Big\rangle \Bigg] . 
	 \label{eq:Wick_decomp}
\end{split}
\end{align}
Here the sum runs over all possible partitions $\pi_2$ of $\{1,\dots,n\}$ into blocks of size 2. The product again runs over all blocks $B$ of the partition. For more details see \cite{gardiner2009stochastic,shiryaev2016probability} and the SI.

\subsection{Integral measure and statistical analysis}
\label{meth:integrated_dev}

To quantify the validity of the Wick decomposition, we use a measure similar to $M^{(N)}$ defined in Eq.~\eqref{eq:int_measure}. We define
\begin{equation}
M_{\mathrm{Wick}}^{(N)}=\frac{\sum_{\bf{z}}\left|G^{(N)}({\bf{z}},0) - G^{(N)}_{\mathrm{Wick}}({\bf{z}},0)\right|}{\sum_{\bf{z}}\left|G^{(N)}({\bf{z}},0)\right|}. \label{eq:wick_int_measure}
\end{equation}
Here $G_{\mathrm{Wick}}^{(N)}({\bf{z}},{\bf{z}}')$ represents the Wick decomposition (see \Eqref{eq:Wick_decomp}).

Calculating the measure $M^{(N)}$ or $M_{\mathrm{Wick}}^{(N)}$  one can significantly reduce the computational effort by utilising the symmetries of the correlation functions under exchange of coordinates. One therefore only has to sum over the correlation function for distinct combinations of $z$-values. Using this symmetry, we still have to reduce the number of points $z$ by a factor 2, considering only every second $z$-value, when calculating the measures for $N = 6$. For $N = 8$ ($N = 10$) only every third (fourth) point was considered. With this, the sum runs over 20475 ($N = 4$), 18564 ($N = 6$), 12870 ($N = 8$), and  8008 ($N = 10$)  terms.

The confidence intervals for the integral measures are calculated by the bootstrap bias corrected and accelerated (BCA) method \cite{efron1986bootstrap} (its implementation 'bootci' in matlab was used). 

\subsection{Sine-Gordon model}
\label{methods:SG}

\begin{figure}[h]
	\begin{center}
		\includegraphics[width=0.7\columnwidth]{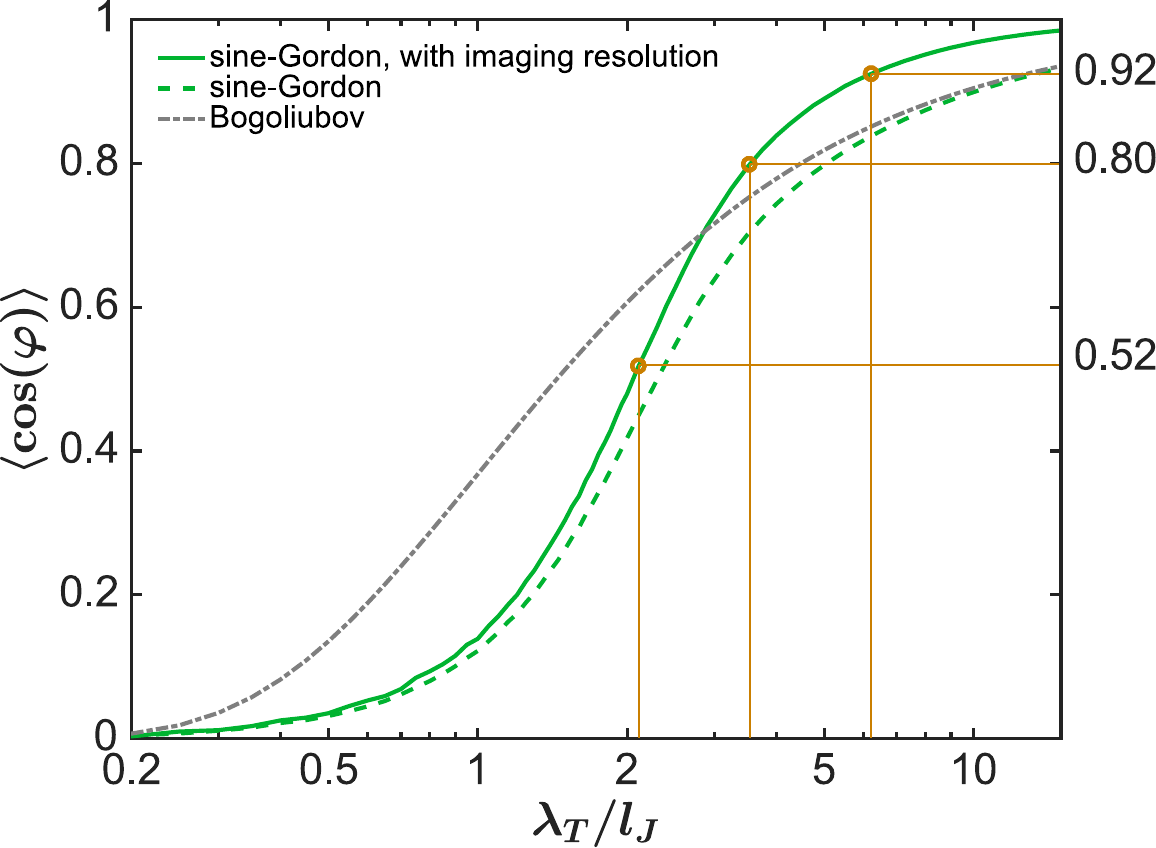}
	\end{center}
	\caption{  { \bf Parameter dependence of phase locking in thermal sine-Gordon theory.} We use the experimental observable $\langle \operatorname{cos}(\phi) \rangle$ to quantify the phase locking in a direct and model independent way. Its dependence on the dimensionless parameter $q = \lambda_T / l_J$ (see Methods \ref{methods:SG}) is displayed for the full sine-Gordon model (dashed green line) and for the quadratic model (dashed-dotted grey line), obtained by expanding the cosine to second order and valid for large phase locking. To compare experiment and theory, the finite imaging resolution needs to be taken into account (solid green line). Some experimental parameters used in the main text are marked and connected to the respective values of $q$. 
	}
	\label{fig:regimes}
\end{figure}

In our parameter regime thermal fluctuations dominate and the sine-Gordon model (see \Eqref{eq:SG}) is characterised by two scales \cite{grivsins2013coherence}: The phase coherence length $\lambda_T=2\hbar^2n_\mathrm{1D}/(mk_BT)$ describing the randomisation of the phase due to the temperature $T$, and the healing length of the relative phase $l_J=\sqrt{\hbar/(4mJ)}$ determining the restoration of the phase coherence through the tunnel-coupling $J$. The dimensionless ratio $q = \lambda_T/l_J$ is directly related to the observable quantity $\braket{\cos(\varphi)}$ and therefore determines the relevance of non-quadratic contributions to the Hamiltonian (see the SI and \EDfigref{fig:regimes}). 

In the experiment, the ratio $q$ can be tuned over a large range by changing the barrier height (corresponds to changing the tunnel-coupling $J$). The phase coherence length was chosen as $\lambda_T \simeq 18\ \mu$m. To be more precise, $\lambda_T = 15 \ldots 20\ \mu$m for the coupled thermal equilibrium data. 
We can measure $\lambda_T$ independently by looking at speckle patterns in time of flight~\cite{manz2010two}. 
Using the same procedure to fit a temperature to the fast cooled (= non-thermal) data gives $\lambda_T = 14 \ldots 27\ \mu$m.

\begin{figure}[h]
\centering\includegraphics[width=0.4\columnwidth]{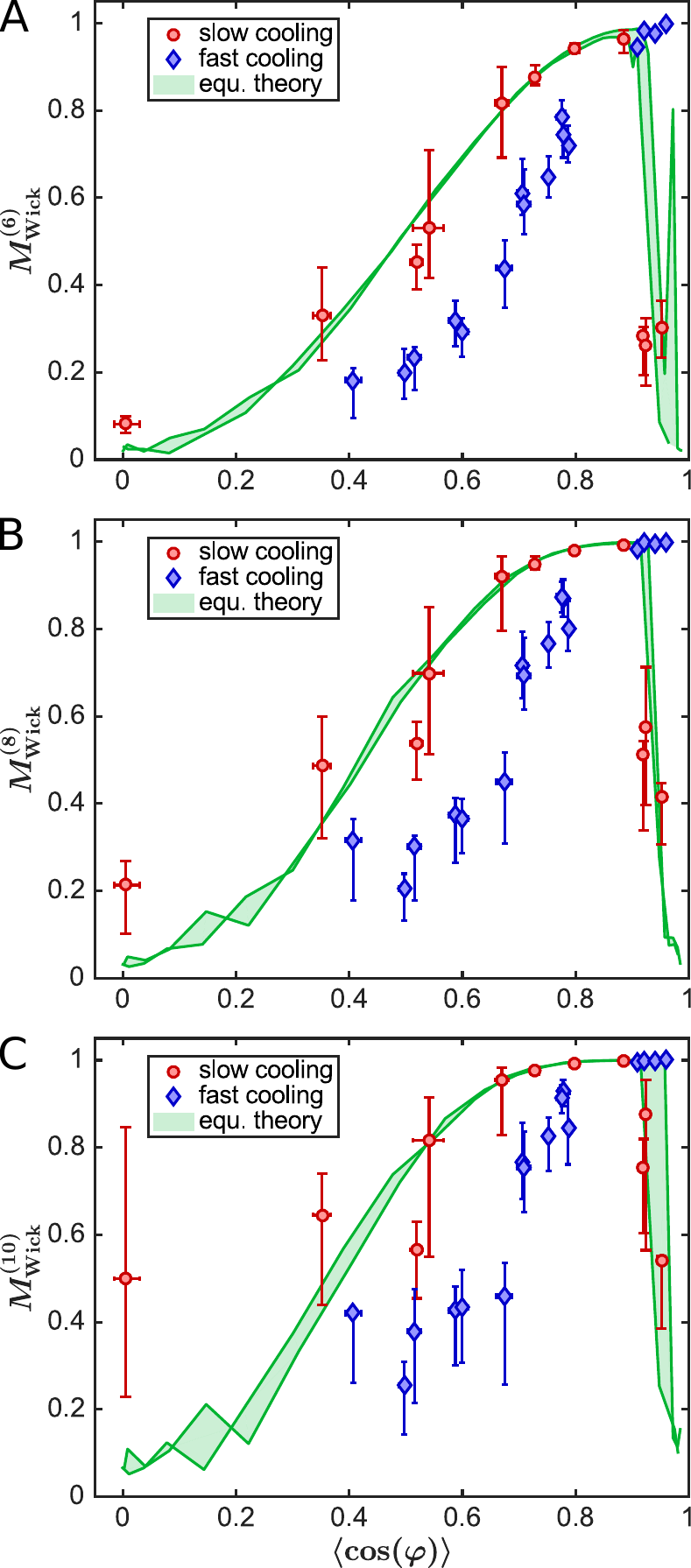}
\caption{ {\bf Relative deviation from the Wick-decomposition,} for the $6^{\mathrm{th}}$- \textbf{(A)}, the $8^{\mathrm{th}}$- \textbf{(B)} and the $10^{\mathrm{th}}$-order \textbf{(C)} correlation function. Plotted is the measure $M_{\mathrm{Wick}}^{(N)}$, as defined in the Methods (\Eqref{eq:wick_int_measure}), vs.\ the phase-locking strength quantified by $\braket{\cos(\varphi)}$. One sees quantitative agreement between thermal equilibrium data (\redbullet) and thermal sine-Gordon theory (\textcolor{plottinggreen}{green shaded region}). The deviation from zero and/or the thermal theory for very small and very big values of $\braket{\cos(\varphi)}$ can be attributed to the finite sample size. Note that for increasing order $N$ increased phase locking is needed to achieve full Wick-factorisation. The fast-cooling data (\bluediamond) clearly deviates from the thermal equilibrium theory prediction. The theory curves were calculated for the maximum spread of the estimated experimental parameters. The error bars represent 80\% confidence intervals calculated by using bootstrapping (see Methods). }
\label{fig:edfigwick6p8p}
\end{figure}

\begin{figure}[h]
\centering\includegraphics[width=0.4\columnwidth]{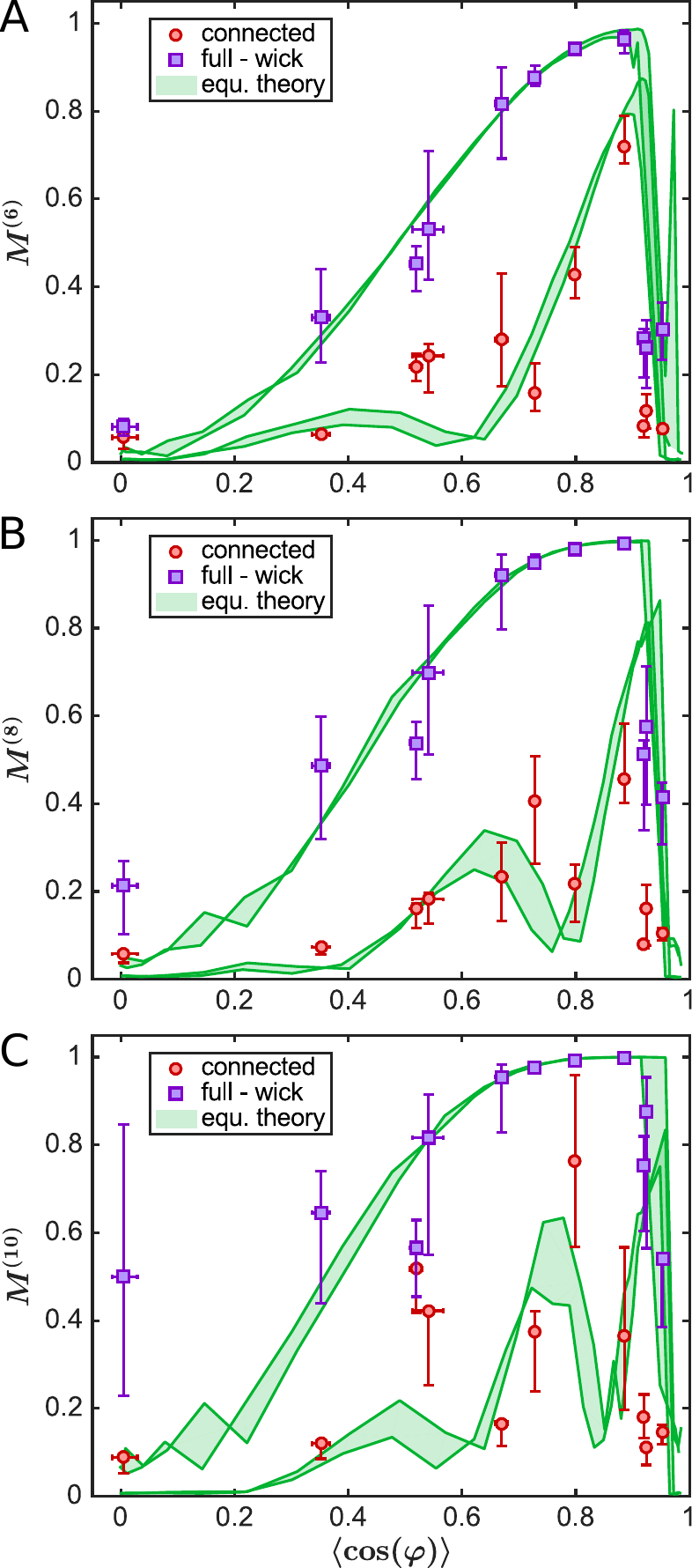}
\caption{ {\bf Relative size of the \boldmath{$6^{\mathrm{th}}$}-, \boldmath{$8^{\mathrm{th}}$}- and \boldmath{$10^{\mathrm{th}}$}-order connected correlation function}, for thermal equilibrium (\redbullet). Plotted is the measure $M^{(N)}$ as defined in \Eqref{eq:int_measure} vs.\ the phase-locking strength quantified by $\braket{\cos(\varphi)}$. One sees qualitative agreement between thermal equilibrium data and thermal sine-Gordon theory (\textcolor{plottinggreen}{green shaded region}). For comparison also the measure $M_{\mathrm{Wick}}^{(N)}$ for the relative deviation from the Wick-decomposition is plotted (\purplesquare). The theory curves were calculated for the maximum spread of the estimated experimental parameters. The error bars represent 80\% confidence intervals calculated by using bootstrapping (see Methods). Note that the apparent deviation from zero and/or the thermal theory for very small and very big values of $\braket{\cos(\varphi)}$ can be attributed to the finite sample size.}	
\label{fig:edfig6pm}
\end{figure}

\begin{figure}
\centering
\includegraphics[width=0.7\linewidth]{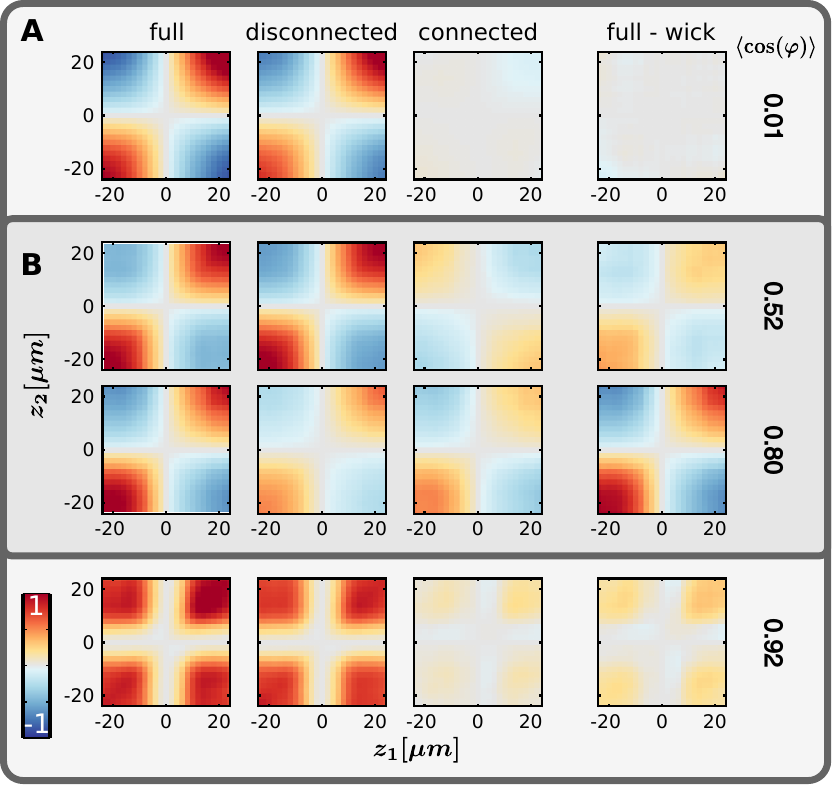}
	\caption{{\bf Decomposition of the \boldmath{$6^{\mathrm{th}}$}-order phase correlation functions $G^{(6)}({\bf{z}},{\bf{z}}')$.} ({\bf A}) uncoupled ($\braket{\cos(\varphi)} \simeq 0$) and strongly phase-locked ($\braket{\cos(\varphi)} \simeq 1$); ({\bf B}) intermediate regime. To visualise the high-dimensional data, we choose $z_3 = -z_4 = 10\ \mu$m, $z_5 = -z_6 = 20\ \mu$m and ${\bf{z}}' = 0$, leading to the observed symmetric crosses where the correlation function vanishes. The colour marks the value of the full, disconnected and connected correlation functions as well as the difference between the full correlation functions and their Wick decompositions. Each row is normalised to its maximum value, implying colour to encode the interval from $-1$ to $1$. While the connected part for $\braket{\cos(\varphi)}=0.92$ is small, the deviation from the Wick decomposition is larger, in agreement with the theory. Full Wick-factorisation requires even higher phase locking (see \EDfigref{fig:edfigwick6p8p}).}
\label{fig:edfig6pslices}
\end{figure}

\begin{figure}
	\centering
	\includegraphics[width=0.7\linewidth]{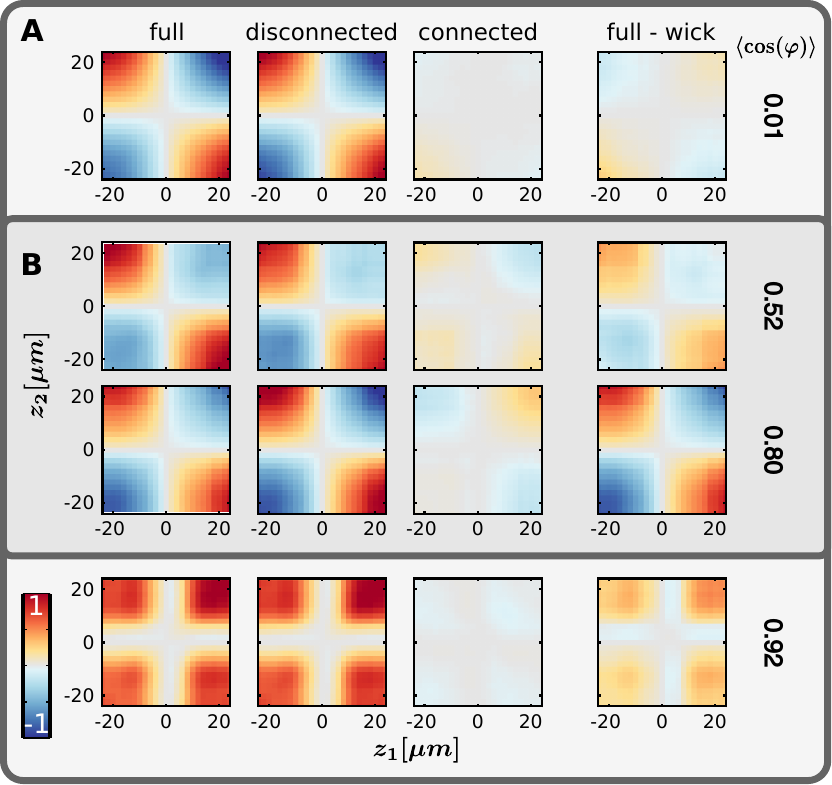}
	\caption{{\bf Decomposition of the \boldmath{$8^{\mathrm{th}}$}-order phase correlation functions $G^{(8)}({\bf{z}},{\bf{z}}')$.} ({\bf A}) uncoupled ($\braket{\cos(\varphi)} \simeq 0$) and strongly phase-locked ($\braket{\cos(\varphi)} \simeq 1$); ({\bf B}) intermediate regime. To visualise the high-dimensional data, we choose $z_3 = -z_4 = 10\ \mu$m, $z_5 = -z_6 = 18\ \mu$m, $z_7 = -z_8 = 24\ \mu$m and ${\bf{z}}' = 0$, leading to the observed symmetric crosses where the correlation function vanishes. The colour marks the value of the full, disconnected and connected correlation functions as well as the difference between the full correlation functions and their Wick decompositions. Each row is normalised to its maximum value, implying colour to encode the interval from $-1$ to $1$. Note that the apparent deviation from zero for the uncoupled case can be attributed to the finite sample size. While the connected part for $\braket{\cos(\varphi)}=0.92$ is small, the deviation from the Wick decomposition is larger, in agreement with the theory. Full Wick-factorisation requires even higher phase locking (see \EDfigref{fig:edfigwick6p8p}).}
	\label{fig:edfig8pslices}
\end{figure}

\begin{figure}
	\centering
	\includegraphics[width=0.7\linewidth]{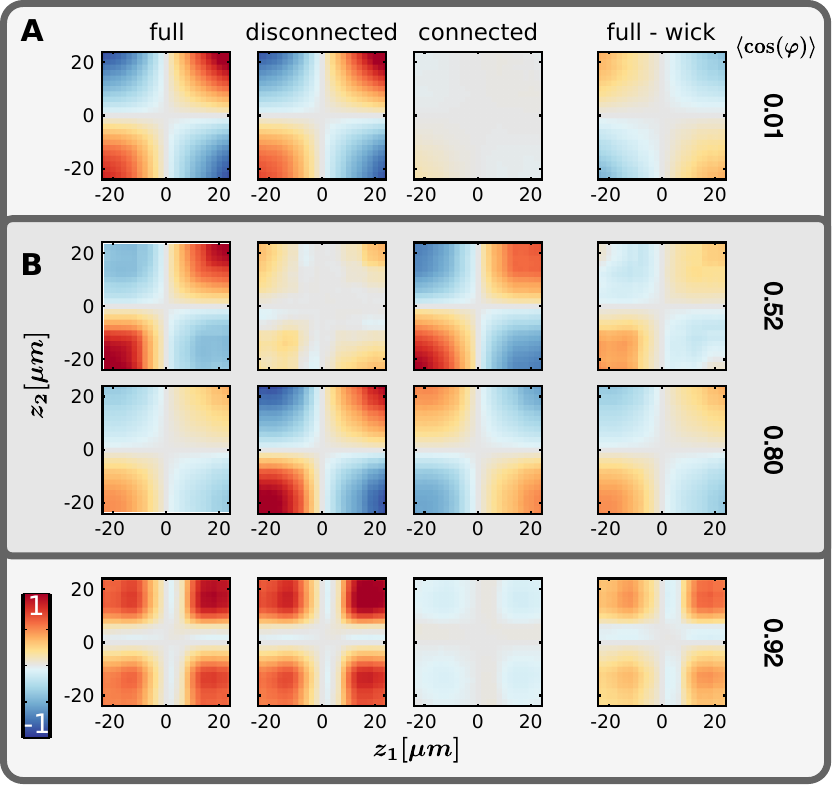}
	\caption{{\bf Decomposition of the \boldmath{$10^{\mathrm{th}}$}-order phase correlation functions $G^{(10)}({\bf{z}},{\bf{z}}')$.} ({\bf A}) uncoupled ($\braket{\cos(\varphi)} \simeq 0$) and strongly phase-locked ($\braket{\cos(\varphi)} \simeq 1$); ({\bf B}) intermediate regime. To visualise the high-dimensional data, we choose $z_3 = -z_4 = 10\ \mu$m, $z_5 = -z_6 = 15\ \mu$m, $z_7 = -z_8 = 20\ \mu$m, $z_9 = -z_{10} = 24\ \mu$m and ${\bf{z}}' = 0$, leading to the observed symmetric crosses where the correlation function vanishes. The colour marks the value of the full, disconnected and connected correlation functions as well as the difference between the full correlation functions and their Wick decompositions. Each row is normalised to its maximum value, implying colour to encode the interval from $-1$ to $1$. 
	Note that the apparent deviation from zero for the uncoupled case can be attributed to the finite sample size. While the connected part for $\braket{\cos(\varphi)}=0.92$ is small, the deviation from the Wick decomposition is larger, in agreement with the theory. Full Wick-factorisation requires even higher phase locking (see \EDfigref{fig:edfigwick6p8p}).}
	\label{fig:edfig10pslices}
\end{figure}

Knowing $\lambda_T$ we can then calculate the measure $M^{(N)}$ as a function of $\braket{\cos(\varphi)}$ by varying the parameter $q$. To compare experiment and theory we plot both datasets in the same figure, yielding \figref{fig:integrated_thermal} as well as \EDfigref{fig:edfigwick6p8p}--\ref{fig:edfig6pm}. The theoretical calculations rely on a random process (see the SI), which is a generalisation of the Ornstein-Uhlenbeck process used for quadratic theories \cite{stimming2010fluctuations}. The theoretical quantities have been calculated from $10^5$ numerical realisations generated by that process. 

The finite imaging resolution was considered by convolving the numerically generated phase profiles with a Gaussian having standard deviation $\sigma = 3\ \mu$m. This value for $\sigma$ was inferred from the coherent transfer function of the imaging system by simulating artificial images and analysing them with the same codes as the experimental absorption images.

\subsection{Non-vanishing higher-order connected correlation functions imply quasiparticle interactions}

The experiment provides information about the \emph{relevance} of the operators of the effective Hamiltonian in describing the dynamics of the relative degrees of freedom (DOF) of two tunnel-coupled superfluids. The relevant operators are obtained by writing the Hamiltonian of the coupled 1D Gross-Pitaevskii systems in density-phase representation of the Bose fields, $\psi_j = \operatorname{exp}(\i \Theta_j) \, \sqrt{n_\mathrm{1D} + \delta\rho_j}$, expanding in powers of the density fluctuations $\delta\rho_j / n_\mathrm{1D}$ and phase gradients $\partial_z \Theta_j$, and transforming to relative and symmetric DOF (see the SI for details of the calculation). In second order, and neglecting mixed density-phase terms, the relative DOF are described by the sine-Gordon Hamiltonian, \Eqref{eq:SG}.

Approximating the cosine to second order in $\varphi$, the Hamiltonian can be diagonalised by means of a Bogoliubov transformation. In the case that all higher-order corrections to this mean-field Hamiltonian represent irrelevant operators, the system's dynamics is determined by a set of non-interacting quasiparticles. As a result, Wick's theorem applied to the quasiparticle creation and annihilation operators $b^{(\dagger)}$ implies the factorisation of $N^{\mathrm{th}}$-order correlation functions into $2^{\mathrm{nd}}$-order functions. The experimental data shows this factorisation for the decoupled as well as for the strongly coupled systems in equilibrium.

In contrast, in the case that terms of higher order in the relative phase and density fluctuations become relevant, the quasiparticles interact with each other and the state no longer shows Gaussian fluctuations. Higher-order connected correlation functions in $b^{(\dagger)}$ can be calculated within a perturbation expansion in the relevant coupling constants. The leading, fourth-order cumulant contains information about the interactions between two quasiparticles, and higher cumulants encode multi-particle correlations. We emphasise that experimentally measured higher-order cumulants give access to the respective information beyond perturbative quantities, cf. the SI for more details.

Compared with the above cumulants for the relative-phase fluctuations, the correlation functions
\begin{align}\label{eq:exp_correlations}
\mathcal{C}(z,z') = \braket{e^{i [\varphi(z)-\varphi(z')]}} \,
\end{align}
are more commonly used in describing low-dimensional superfluids, e.g., in a Luttinger-liquid formulation. However, they encode the fluctuations of the underlying Bose fields in the limit where density fluctuations are negligible. Already the $2^{\mathrm{nd}}$-order correlation function contains information about the interactions to arbitrary order in the phase-fluctuation fields. For our system of two tunnel-coupled superfluids, this renders these correlations not suitable for the general method presented here of analysing many-body problems by higher-order correlation functions.

\onecolumngrid

\renewcommand{\thesection}{}
\titlelabel{}
\section{Extended Data}
\titlelabel{\thetitle.}

\clearpage    
\appendix

\centerline{ \Large{\textbf{Supplementary Information}} }
%
\renewcommand{\theequation}{S\arabic{equation}}



\vspace*{1.5cm}
These supplementary informations give a brief overview of the theoretical calculations and concepts used in the main text. In the first part the low-energy 
effective theories for two coupled one-dimensional superfluids and their exact solution within the classical-field approximation are discussed. In the second part, 
we first provide a general introduction to equal-time correlation functions. We furthermore give explicit formulas for the experimentally 
measured correlation functions and their decomposition into connected and disconnected parts, and discuss the connection to periodic observables used in our previous publications.
The last section discusses the connection of the measured correlation functions and quasiparticle interactions.

\section{Theoretical models}

In the first section, we introduce the one-dimensional model of two tunnel-coupled superfluids and discuss the derivation of the sine-Gordon Hamiltonian, proposed as a low-energy effective theory for the relative degrees of freedom.Thereafter, in the second section, we discuss the exact solution within the classical-field approximation, using the transfer matrix formalism.

\subsection{Sine-Gordon model as effective low-energy theory for two tunnel-coupled superfluids}
\label{sec:sGlEeffTh}
The quantum many-body system we study is an ultracold gas of $^{87}\mathrm{Rb}$ atoms in a double-well (DW) potential on an atom chip as shown in Fig. 1 in the main paper. 
Each well is tightly confined in the radial direction ($\omega_{\perp} \simeq 2 \pi \times 1.4 \, \mathrm{kHz}$) and weakly confined along the 
longitudinal direction ($\omega_z \simeq 2 \pi \times 6.7 \, \mathrm{Hz}$).
Since both superfluids fulfill the condition of being one-dimensional (1D), $\mu, k_{\mathrm{B}}T < \hbar \omega_{\perp}$, dynamics along the radial direction is frozen. However, tunneling through the adjustable DW 
barrier couples the two superfluids. Integrating over the radial degrees of freedom, reduces the problem to an effective 1D system
described by the Hamiltonian
\begin{align} \label{eq:Hcoupledcondensates}
H = &\sum_{j=1}^2 \int dz  \left[
\frac{\hbar^2}{2m} \frac{\partial {\psi}^{\dagger}_j }{\partial z }  \frac{\partial {\psi}_j }{\partial z } 
 +\frac{g_{\mathrm{1D}}}{2}  {\psi}^{\dagger}_j {\psi}^{\dagger}_j {\psi}_j {\psi}_j +\, {U}(z) \psi^{\dagger}_j \psi_j 
- \mu {\psi}_j^{\dagger} {\psi}_j \right] \notag \\
&- \hbar J \int dz \left[ {\psi}^{\dagger}_1 {\psi}_2 + {\psi}^{\dagger}_2 {\psi}_1\right] \, .
\end{align}
Here $m$ is the atomic mass, $g_{\mathrm{1D}} = 2 \hbar a_\mathrm{s} \omega_{\perp}$ the 1D effective
interaction strength, calculated from the $s$-wave scattering length $a_\mathrm{s}$ and the frequency $\omega_{\perp}$ of the radial confinement. 
$U(z)$ is  the trapping potential in the longitudinal direction, and $\mu$ the chemical potential. 
The field operators fulfill the bosonic commutation relations $[\psi_j(z), \psi^{\dagger}_{j'}(z')] = \delta_{j j'} \delta(z-z')$. 
For simplicity, we consider in the following the homogenous case, $U(z) \equiv 0$. In order to derive a low-energy effective theory we use 
the density-phase representation,
\begin{align}
{\psi}_j(z) = \exp [i {\theta}_j (z)] \sqrt{n_\mathrm{1D} + \delta {\rho}_j(z)} ~\mathrm{,}
\label{eq:DensityPhase}
\end{align}
with the canonical commutators $[\delta {\rho}_j(z), \theta_{j'}(z')] = \i \, \delta_{j j'}\delta(z-z')$.
We define the symmetric (subscript $s$) and anti-symmetric (subscript $a$) degrees of freedom as
\begin{align}
&\delta {\rho}_s(z) = \delta {\rho}_1(z) + \delta {\rho}_2(z)  \hspace{0.01\textwidth}~\mathrm{,} &&{\varphi}_s(z) = \frac{1}{2} [{\theta}_1(z) + {\theta}_2(z)]~\mathrm{,} \\
&\delta {\rho}_a(z) = \frac{1}{2} [\delta {\rho}_1(z) - \delta {\rho}_2(z)]~\mathrm{,} \hspace{0.01\textwidth} &&{\varphi}_a(z) = {\theta}_1(z) - {\theta}_2(z) ~\mathrm{.}
\end{align}
Evidently these fields fulfil canonical commutation relations as well. Expanding the Hamiltonian \eqref{eq:Hcoupledcondensates} in powers of the density fluctuations $\delta \rho_j$
and phase gradients $\partial_z \varphi_j$ to quadratic order leads to
\begin{align}
H = H_s[\delta \rho_s,\varphi_s] 
&+ \int dz \biggl[
\frac{\hbar^2}{4m n_\mathrm{1D}} \biggl( \frac{\partial {\delta \rho_a}}{\partial z}\biggr)^2 + g_{\mathrm{1D}}\delta {\rho_a}^2 + 
\frac{\hbar^2 n_\mathrm{1D}}{4m} \biggl( \frac{\partial {\varphi_a}}{\partial z}\biggr)^2  - 2\hbar J n_\mathrm{1D} \cos \varphi_a  \biggr] \notag \\
&+ \int dz ~ \biggl[ \frac{\hbar J}{n_{\mathrm{1D}}} \delta \rho_{a} ( \cos \varphi_a ) \delta \rho_{a} - \hbar J \, \delta\rho_s \cos \varphi_a \biggr] \label{eq:fullSecondOrder_densphasecoupling} \, ,
\end{align}
where the Hamiltonian $H_s$ depends only on the symmetric degrees of freedom. Note that, while phase gradients are expected to be small for all values of $J$, 
the phase field $\varphi$ itself needs to be considered non-perturbatively, leading to the full cosine potentials. The last term couples 
the symmetric and antisymmetric degrees of freedom and is expected to be significant for, e.g., the non-linear relaxation of the system following a quantum-quench. 
In, or close to, thermal equilibrium it is presumed that the couplings of density and phase fluctuations are negligible, which leads to a complete 
decoupling of the symmetric and antisymmetric degrees of freedom.
The low-energy effective Hamiltonian describing the relative degrees of freedom takes the form
\begin{align}
H = \int dz \biggl[ 
\frac{\hbar^2}{4m n_\mathrm{1D}} \biggl( \frac{\partial {\delta \rho}}{\partial z}\biggr)^2  + g\delta {\rho}^2 + 
\frac{\hbar^2 n_\mathrm{1D}}{4m} \biggl( \frac{\partial {\varphi}}{\partial z}\biggr)^2 -\, 2\hbar J n_\mathrm{1D} \cos \varphi \biggr] \label{eq:fullSecondOrder} \, ,
\end{align}
where we introduced $g = g_{\mathrm{1D}} + \hbar J/n_\mathrm{1D}$ and omitted the subscript $a$, as we do in the following, and in the main text. 
For $J = 0$, the Hamiltonian reduces to
\begin{align}
H \!\!=\!\! \int dz \biggl[ \frac{\hbar^2}{4m n_\mathrm{1D}} \biggl( \frac{\partial {\delta \rho}}{\partial z}\biggr)^2 + 
g \delta {\rho}^2 + \frac{\hbar^2 n_\mathrm{1D}}{4m} \biggl( 
\frac{\partial {\varphi}}{\partial z}\biggr)^2
\biggr] \label{eq:fullSecondOrderDecoupled} .
\end{align}
On the other hand, for $\langle \cos(\varphi) \rangle \approx 1$, i.e. for strong 
tunnel-coupling $J$, fluctuations of the phase are tightly centered around zero, and the cosine in \Eq{fullSecondOrder} can be expanded to 
second order leading to
\begin{align}
H \!= \int \! \!dz \! &\left[
\frac{\hbar^2}{4m n_\mathrm{1D}} \left( \frac{\partial {\delta \rho}}{\partial z}\right)^2 + g  \delta {\rho}^2
 + \frac{\hbar^2 n_\mathrm{1D}}{4m} \left( 
\frac{\partial {\varphi}}{\partial z}\right)^2  \vphantom{\left(g+ \frac{\hbar J}{n_\mathrm{1D}}\right)^2} +   \hbar J n_\mathrm{1D} {\varphi}^2
\right] \label{eq:fullSecondOrderStrongcoupled} \!.
\end{align}

Both Hamiltonians, \Eq{fullSecondOrderDecoupled} and \Eq{fullSecondOrderStrongcoupled}, are quadratic in the fields and can therefore be 
diagonalized by a Bogoliubov transformation (see below). The system is described by 
non-interacting quasi-particles with a gap proportional to $\sqrt{J}$. 
Note that this remains valid for a non-vanishing external potential $U(z)$, although the explicit form of the dispersion relation changes.

At the low energies considered, density fluctuations are highly suppressed and hence one can neglect the term involving the derivative of the relative density, thereby restricting the spectrum to the phononic regime. 
For the uncoupled system, \Eq{fullSecondOrderDecoupled}, this leads to the Tomonaga-Luttinger Hamiltonian, 
whereas \Eq{fullSecondOrder}, valid for general couplings $J$, reduces to the sine-Gordon Hamiltonian,
\begin{align}
H_{\mathrm{SG}} = \int dz \left[ g \delta {\rho}^2 + 
\frac{\hbar^2 n_\mathrm{1D}}{4m} \left( \frac{\partial {\varphi}}{\partial z}\right)^2  -\, 2\hbar J n_\mathrm{1D} \cos \varphi 
\right] \label{eq:SGextended} \, .
\end{align}

At the classical level, the equations of motion derived from this Hamiltonian include solitonic and breather solutions. The single 
soliton/anti-soliton solution is given by
\begin{align}
\varphi_{\mathrm{S}}(z) =  4 \arctan \left[ \pm \exp \frac{z  - z_0 - v_{\mathrm{S}} t}{l _J \sqrt{1-(v_{\mathrm{S}}/c_\mathrm{s})^2}} \right] \, ,
\end{align}
where $z_0$ is the position and $v_{\mathrm{S}}$ the velocity of the soliton, and $c_{\mathrm{s}}=\sqrt{gn_{\mathrm{1D}}/m}$ the speed of 
sound (see, \textit{e.g.}\ \cite{cuevas2014sine}). The width of the soliton is given by the length scale $ l_J = \sqrt{\hbar / 4mJ}$. 
Motion of the soliton leads to a contraction of this length scale by the `Lorentz' factor $\sqrt{1-(v_{\mathrm{S}}/c_\mathrm{s})^2}$.
These topological defects represent a local phase-twist of $2 \pi$, connecting adjacent minima of the cosine potential.

The sine-Gordon model represents an exactly-solvable field theory. The sine-Gordon Hamiltonian \eqref{eq:SGextended} can be written 
in the re-scaled form
\begin{align} \label{eq:SGtheory}
 H_{\mathrm{SG}} = \frac{1}{2} \int dz \left[ \Pi^2 + (\partial_z \phi)^2 - \Delta \cos{\beta \phi} \right] \, \mathrm{,}
\end{align}
where we set $\hbar=k_{\mathrm{B}}=1$, rescaled time $t \to c_s t$, and set $c_s = 1$. 
Furthermore, we define the conjugate momentum $\Pi = \beta \, \delta \rho$, the rescaled phase field $\phi = \varphi / \beta$, 
as well as the parameters $\beta = \sqrt{2 \pi / K}$ and $\Delta = 8 J m / \beta^2$. 
The Luttinger parameter $K$ is, in the weakly interacting regime ($\gamma \ll 1$), given by $K \approx \pi / \sqrt{\gamma}$, where
$\gamma = m g / \hbar^2 n_{\mathrm{1D}}$, characterising the strength of the interaction.
For theoretical studies of the sine-Gordon model see \textit{e.g.}\ \cite{Gritsev07, Bertini2014}. The parameters applying to our experiment correspond to 
the weakly interacting regime, $K \gg 1$, typically $K=63 \dots 73$, and hence $\beta^2=0.1 \dots 0.086$.

For completeness, we give a brief overview of the different regimes of the sine-Gordon model, supposing $\beta$ and $\Delta$ as independent parameters. 
The spectrum of the Hamiltonian, \Eq{SGtheory}, depends on the value of $\beta$. The system undergoes a Kosterlitz-Thouless 
transition at the critical point $\beta^2 = 8 \pi$. For larger values, $\beta^2>8 \pi$, the cosine term becomes irrelevant and the system reduces to 
the Luttinger-Liquid model. As was shown in \cite{Coleman75} for $\beta^2<8 \pi$ the sine-Gordon model is equivalent to the zero-charge
sector of the massive Thirring model, describing massive Dirac fermions with local self-interaction. In this regime, the spectrum can be further 
divided into two distinct sectors, separated by the Luther-Emery point, $\beta^2=4 \pi$, at which the model describes non-interacting massive Dirac fermions. For 
$4 \pi < \beta^2 < 8 \pi$, the system is described by soliton and anti-soliton excitations, whereas for $0 < \beta^2 < 4 \pi$, the spectrum contains 
additional bound states of (anti-)solitons, called \textit{breathers}. 

\subsection{Exact results within the classical-field approximation}

Within the classical-field approximation, correlation functions of the system at temperature $T$ can be calculated using the transfer-matrix 
formalism developed in \cite{Krum1,Krum2}. The harmonic approximation \eq{fullSecondOrderStrongcoupled} for two tunnel-coupled superfluids has been analysed in \cite{stimming2010fluctuations}. 
In particular, the Gaussian fluctuations of the phase along $z$ have been shown to be describable by an Ornstein-Uhlenbeck process.\footnotemark[1] 
This enables the efficent sampling of the fields, directly from the equilibrium distribution. Here we sketch the 
extension of these methods for the case of two coupled wave guides described by \Eq{Hcoupledcondensates} beyond the harmonic approximation. 
\footnotetext[1] {Note that we deal with stochastic processes evolving in space, along $z$, but not in time.}

The system realized in our experiments is a special case of the model described by the (classical) Hamiltonian
\begin{eqnarray} 
H=\int dz\, \Big{[ } \sum _{j =1}^{M} \left( \frac {\hbar ^2}{2m} \frac {\partial \psi_j ^*}{dz} 
\frac {\partial \psi_j }{dz} -\mu \psi_j ^* \psi_j \right) 
+ V(\psi_{M}^*, \, \dots \, ,\psi_1^*,\psi_1, \, \dots \, ,\psi_{M}) \Big{ ] }, 
\label{H-gen} 
\end{eqnarray}
for the $M$-component Bose field $\psi _j (z)$, $j=1,\dots,M$, with an arbitrary local, but not necessarily pairwise interaction potential $V$, conserving the total number of atoms, 
$N=\int dz\, \sum _{j =1}^{M} \psi_j ^*\psi_j $ (I. Mazets, in preparation). Comparing with \Eq{Hcoupledcondensates} we get ($M = 2$)
\begin{align}
 V = \frac{g}{2}  \left[ ({\psi}^*_1 {\psi}_1)^2 + ({\psi}^*_2 {\psi}_2)^2 \right]
     - \hbar J \left[ {\psi}^*_1 {\psi}_2 + {\psi}_2 {\psi}^*_1\right] \, ,
\end{align}
and the chemical potential $\mu = g n_{1D} - \hbar J$.

The transfer-matrix formalism \cite{Krum1,Krum2} yields the following expressions for the thermal average and correlation function of operators 
$\mathcal{O}(z)$: 
\begin{eqnarray} 
\langle \mathcal{O}(z_1)  \rangle& =&\sum _n \langle 0|\mathcal{O}(z_1)|0\rangle , \label{eq:corr-1} \\ 
\langle \mathcal{O}(z_1) \mathcal{O}(z_2) \rangle& =&\sum _n \langle 0|\mathcal{O}(z_2)|n\rangle \langle n|\mathcal{O}(z_1)|0\rangle 
e^{-(\kappa _n-\kappa _0)(z_2-z_1)} \qquad (z_2\geq z_1) ,
\label{eq:corr-2} 
\end{eqnarray}  
where the matrix elements with respect to the eigenstates $|n\rangle$ of the transfer operator $\hat{K}$ (see below), with eigenvalues $\kappa_n$, 
are defined as: 
\begin{eqnarray} 
\langle n^\prime |\mathcal{O}(z)|n\rangle & = &\int_{0}^{\infty} dr_1 \, r_1 \int_{0}^{\infty} dr_2 \, r_2 \int_{0}^{2 \pi} d\theta_2 
\int_{0}^{2 \pi} d\theta_2 \, \Psi_{n^\prime }^* \mathcal{O}(z)\Psi_n \, . ~~
\label{matr-elem} 
\end{eqnarray} 
Here, we introduced the density $r_j^2$, and the phase $\theta_j$ is defined, in analogy to the previous discussion, via 
$\operatorname{Re}(\psi_j) = r_j \cos \theta_j$ and $\operatorname{Im}(\psi_j) = r_j \sin \theta_j$. The observables  
$\mathcal{O}(z)=\mathcal{O}(r_1, \theta_1, r_2 ,\theta_2)\vert_z$ can be arbitrary functions of the classical field provided the integrals exist. 
The eigenvalues $\kappa_n$ and orthonormal eigenfunctions $\Psi_n = \Psi_n(r_1, \theta_1, r_2 ,\theta_2)$ 
are given by the Hamiltonian-like hermitian operator $\hat K$ that arises in the transfer matrix formalism \cite{Krum1,Krum2}. 
For our system of two tunnel-coupled superfluids we have
\begin{align}
 \hat K=\hat K_1^\mathrm{s}+\hat K_2^\mathrm{s} +\frac {\hbar J}{k_\mathrm{B}T} (r_1^2+r_2^2)
- \frac {2\hbar J}{k_\mathrm{B}T}r_1r_2\cos (\theta _1 -\theta _2) , 
\label{eq:K-2q} 
\end{align}
where 
\begin{align} 
\hat K_j ^\mathrm{s} = -D \left( \frac 1{r_j }\frac \partial {\partial r_j }r_j 
\frac \partial {\partial r_j }+\frac 1{r_j^2}\frac {\partial^2}{\partial \theta_j^2}\right) 
+\frac g{2k_\mathrm{B} T} {r_j^2}({r_j^2}-2n_\mathrm{1D}) 
\label{eq:K-1q} 
\end{align}
is the auxiliary operator for a single superfluid, and $D=mk_\mathrm{B}T/(2\hbar ^2)$. The equilibrium distribution of the real classical 
variables is determined by the ground (lowest-eigenvalue) state of the operator $\hat K$ \cite{Krum1,Krum2}, via
\begin{align}
 W_{\mathrm{eq}}(r_1,\theta_1,r_2,\theta_2) = |\Psi_0(r_1,\theta_1,r_2,\theta_2)|^2 \, .
\end{align}
It is possible to construct a Fokker-Planck equation for the classical probability distribution $W(r_1,\theta_1,r_2,\theta_2;z)$ that describes the same stochastic process as the transfer-matrix formalism:
\begin{align}
 \frac \partial {\partial z}W = \sum_{j=1}^{2 N_f}\left[ D \frac {\partial ^2}{\partial q_j^2}W+ 
\frac \partial {\partial q_j} (A_{q_j} W )\right] ~\mathrm{.}
\label{eq:FP-gen} 
\end{align}
To shorten the notation, the variables 
$\operatorname{Re}(\Psi_{1,2})$ and $\operatorname{Im}(\Psi_{1,2})$ are renamed as $q_j \, (j=1,2,3,4)$. The stationarity condition of the equilibrium solution $\partial_z W_{\mathrm{eq}} = 0$ 
determines the drift coefficients $A_{q_j}$, for which we obtain from \Eq{FP-gen}:
\begin{align} \label{eq:drifcoef}
 A_{q_j} \equiv A_{q_j}(q_1,q_2,q_3,q_4) = - D \frac \partial {\partial q_j} \ln W_\mathrm{eq} = -2D \frac \partial {\partial q_j} \ln |\Psi _0| \, .
\end{align}
The last step is to realise, that the Focker-Planck equation is equivalent to a stochastic process described by an Ito equation \cite{gardiner2009stochastic}
\begin{equation}
 dq_j =-A_{q_j}\, dz +\sqrt{2D}\, dX_z \, , 
\label{eq:Ito-1} 
\end{equation}
where $dX_z$ is a random term obeying Gaussian statistics with zero mean, $\langle dX_z \rangle = 0$, and variance, $\langle dX^2_z \rangle = dz$.
Fast sampling of the fields from the full classical equilibrium probability distribution is possible using \Eq{drifcoef} and \Eq{Ito-1},
after finding only the ground state $\Psi_0$ of the auxiliary operator \eqref{eq:K-2q} instead of the whole spectrum as \Eq{corr-2} requires.
Note that the transfer-matrix formalism provides results for the correlation statistics of the unbound phase difference 
in the limit dominated by thermal fluctuations. This allows us to analyse continuous, unbound phase 
differences $\Delta \varphi$. Arbitrary correlation functions can therefore easily be calculated
by averaging over independently sampled field configurations. 

In the limit of vanishing tunnel coupling $J$, we obtain $A_{\varphi} \equiv 0$, 
\textit{i.e.}, the relative phase is described by a diffusion process. In the opposite limit of strong tunnel coupling, we recover the results 
of Ref.~\cite{stimming2010fluctuations}. The sine-Gordon Hamiltonian, \Eq{SGextended}, relevant for intermediate $J$, is described by the auxiliary operator $\hat K$ 
given in \Eq{K-2q}, neglecting the non-linear coupling between fluctuations of the relative phase $\varphi = \theta_1 - \theta_2$ and the 
densities $r^2_{1,2}$. In this approximation, the symmetric phase and the densities are determined by the usual 
Gaussian diffusion and Ornstein-Uhlenbeck processes, respectively. The relative phase needs to be calculated by means of the anharmonic model 
\eqref{eq:Ito-1}, $\Psi_0$ being the lowest-eigenvalue solution of the corresponding Mathieu equation \cite{grivsins2013coherence}. 

We compared the results of the direct calculations of the $4^\mathrm{th}$ moment, using Eq.~(\ref{matr-elem}) after diagonalising $\hat K$, to correlation functions obtained by averaging independently 
sampled field configurations and found perfect agreement. To explain the experimental observations we consider the latter approach, as it 
allows to incorporate the finite imaging resolution (see Methods). We furthermore compared the analytical results for the homogenous system to simulations of the 
stochastic Gross-Pitaevskii equation for harmonically trapped tunnel-coupled superfluids.
Thereby, we found good agreement of the correlation functions in the 
central part of the cloud, for the range of experimental parameters.

\section{Correlation functions}

In this part, we first give a general intorduction to equal-time correlation functions, their role in quantum field theory, and their connection 
to the interaction properties of the system. We further give explicit expressions for the correlation functions used in the experiment 
and their decomposition into connected and disconnected parts. Thereafter, we discuss their relation to commonly used periodic correlation functions, 
explaining in detail as to why they are, in general, not suitable to study the interaction properties of our system.
In the last section we discuss, how a perturbative approach to the sine-Gordon model readily reveals the connection between equal-time correlation functions 
of the phase field and $N$-body quasiparticle interactions.

\subsection{Equal-time correlation functions and their relevance in (quantum) field theory} \label{sec:equal-time-corr}

For a quantum many-body system which is described in terms of a Heisenberg field operator $\mathcal{O}(t,x)$, all physical information is contained 
in correlation functions like
\begin{equation}
\langle \mathcal{O}(t_1,x_1) \mathcal{O}(t_2,x_2) \cdots \mathcal{O}(t_N,x_N) \rangle \equiv {\rm Tr} \left\{ \rho_D \, \hat{T}\, \mathcal{O}(t_1,x_1) \mathcal{O}(t_2,x_2) \cdots \mathcal{O}(t_N,x_N) \right\},
\label{eq:correlationdef}
\end{equation}
where we consider, for the moment, a real scalar field with a single component. Here $\rho_D$ denotes the density operator specifying the system 
at a given time, and the trace is taken over the time-ordered product of field operators as indicated by the time-ordering operator $\hat{T}$.
For instance, the $2^{\mathrm{nd}}$-order function
\begin{equation}
G^{(2)}(t_1,x_1;t_2,x_2) \equiv \langle \mathcal{O}(t_1,x_1) \mathcal{O}(t_2,x_2) \rangle 
\end{equation}
quantifies the correlation between the point $x_1$ at time $t_1$ and the point $x_2$ at time $t_2$. 
In the following we assume a vanishing field expectation value, $\langle \mathcal{O}(t,x) \rangle = 0$, as well as vanishing correlations $G^{(N)}$
for odd-integer $N$. In this case, for a non-zero $4^{\mathrm{th}}$-order correlation 
\begin{equation}
G^{(4)}(t_1,x_1;t_2,x_2;t_3,x_3;t_4,x_4) \equiv \langle \mathcal{O}(t_1,x_1) \mathcal{O}(t_2,x_2) \mathcal{O}(t_3,x_3) \mathcal{O}(t_4,x_4) \rangle ~\mathrm{,}
\end{equation}
one can identify the following contributions:
\begin{eqnarray}
G^{(4)}(t_1,x_1;t_2,x_2;t_3,x_3;t_4,x_4) & = &
G^{(4)}_{\mathcal{O},{\rm con}}(t_1,x_1;t_2,x_2;t_3,x_3;t_4,x_4)
\nonumber\\	
&+& G^{(2)}(t_1,x_1;t_2,x_2)\, G^{(2)}(t_3,x_3;t_4,x_4)
\nonumber\\
&+& G^{(2)}(t_1,x_1;t_3,x_3)\, G^{(2)}(t_2,x_2;t_4,x_4)
\nonumber\\
&+& G^{(2)}(t_1,x_1;t_4,x_4)\, G^{(2)}(t_2,x_2;t_3,x_3) \, .
\end{eqnarray}	
Here, the connected $4^{\mathrm{th}}$-order correlation, 
$G^{(4)}_{{\rm con}}$, is obtained from the full correlation by subtracting products of $2^{\mathrm{nd}}$-order correlations. In this way,
the redundant information that is contained in disconnected lower-order correlations is being eliminated. For any higher $N^{\mathrm{th}}$-order correlations, 
with $N \geq 6$, a similar decomposition into connected and disconnected parts exists, the latter involving products of $G^{(2)}$,
$G^{(4)}_{{\rm con}}$, $\ldots$, $G^{(N-2)}_{{\rm con}}$ that are correspondingly defined. Knowing all connected correlation functions 
is equivalent to knowing all full correlation functions and therefore sufficient for recovering all information about the system.

If the density operator $\rho_D$ describes thermal equilibrium, then the correlation functions become time-translation invariant. In this case, 
employing a Fourier transformation with respect to times, one can represent the $N^{\mathrm{th}}$-order correlation (\ref{eq:correlationdef}) as   
\begin{eqnarray}
\langle \mathcal{O}(t_1,x_1) \cdots \mathcal{O}(t_N,x_N) \rangle =
\int \frac{d\omega_1}{2\pi} \cdots
\frac{d\omega_N}{2\pi}\,
e^{i(\omega_1 t_1 + \cdots + \omega_N t_N)}\, 2 \pi\, \delta(\omega_1+\cdots+\omega_N) 
&& \nonumber\\
\qquad \qquad G^{(N)}(\omega_1,\ldots,\omega_{N-1};x_1,\ldots,x_N) \, .
&& \label{eq:amplitudes}
\end{eqnarray}
Here $G^{(N)}(\omega_1,\ldots,\omega_{N-1};x_1,\ldots,x_N)$ denotes the $N^{\mathrm{th}}$-order correlation amplitude with external frequencies 
$\omega_i$ at spatial points $x_i$, for $i=1,\ldots,N$. Diagrammatically, they can be represented as:
\begin{figure}[h!]
	\centering
	\includegraphics[width=0.35\columnwidth]{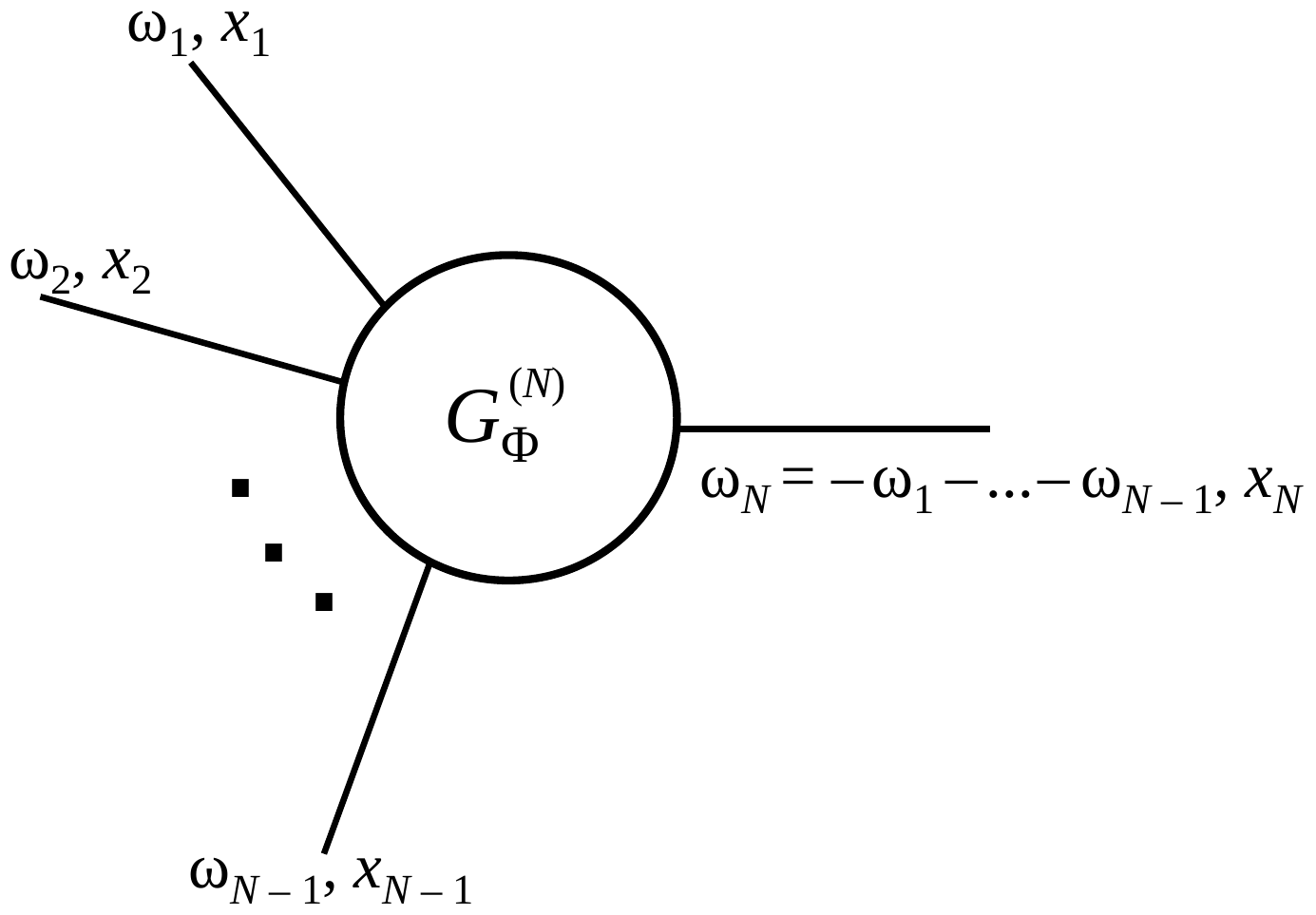}
	\label{fig:Gcon_visual}
\end{figure}
\newline
For instance, the $4^{\mathrm{th}}$-order amplitude $G^{(4)}(\omega_1,\omega_2,\omega_3;x_1,x_2,x_3,x_4)$ describes all possible quantum processes with 
$|\omega_i|$ injected ($\omega_i > 0$) or taken out ($\omega_i < 0$) at points $x_i$ for $i=1,2,3$ such that the total energy is conserved with 
$-\omega_1-\omega_2-\omega_3$ at $x_4$. For two-body interactions and the case of a real scalar field, this will involve standard scattering processes with Feynman diagrams having 
two incoming and two outgoing lines, but also diagrams with one line in and three lines out, the conjugate process (three in, one out), or 
even all lines in (or all lines out).

\footnotetext[2]{
	More precisely, equal-time correlation functions for bosonic fields measure the symmetrized (anti-commutator) part of the time-ordered 
	correlators (\ref{eq:correlationdef}). For the real scalar field operator considered, this can be directly observed from the definition of the 
	time-ordering operator, as e.g.\ for the $2^{\mathrm{nd}}$-order correlation:
	\begin{align*}
	\langle \hat{T} \mathcal{O}(t_1,x_1) \mathcal{O}(t_2,x_2) \rangle
	&= \langle \mathcal{O}(t_1,x_1) \mathcal{O}(t_2,x_2) \rangle \Theta(t_1 - t_2) \\ &+ \langle \mathcal{O}(t_2,x_2) \mathcal{O}(t_1,x_1) \rangle \Theta(t_2 - t_1)	
	\notag\\ 
	&= \frac{1}{2} \langle \left\{ \mathcal{O}(t_1,x_1), \mathcal{O}(t_2,x_2) \right\}\rangle  \\ &+ \frac{1}{2} \langle\left[\mathcal{O}(t_1,x_1), \mathcal{O}(t_2,x_2) \right]\rangle\, {\rm sgn}(t_1-t_2) 
	\notag\, .
	\end{align*}
	Here the step function is defined by $\Theta(t>0)=1$ and $\Theta(t<0) = 0$ and the sign function is ${\rm sgn}(t) \equiv  \Theta(t) - \Theta(-t)$. Since the 
	equal-time commutator vanishes, $[\mathcal{O}(t,x_1), \mathcal{O}(t,x_2)]=0$ for the real scalar field operator, the symmetrized part is given by the 
	anti-commutator $\{ \mathcal{O}(t_1,x_1), \mathcal{O}(t_2,x_2) \}\equiv \mathcal{O}(t_1,x_1) \mathcal{O}(t_2,x_2) + \mathcal{O}(t_2,x_2) \mathcal{O}(t_1,x_1)$ at equal times 
	$t_1 = t_2$. 
}

In this work, we measure equal-time correlation functions, where $t=t_1=t_2= \cdots = t_N$. From the Fourier representation (\ref{eq:amplitudes}) 
one observes that an $N^{\mathrm{th}}$-order equal-time correlation function represents the sum over all the different processes with $N$ external lines%
\footnotemark[2]
\begin{equation}
\langle \mathcal{O}(t,x_1) \cdots \mathcal{O}(t,x_N) \rangle =
\int \frac{d\omega_1}{2\pi} \cdots
\frac{d\omega_{N-1}}{2\pi}\,
 G^{(N)}(\omega_1,\ldots,\omega_{N-1};x_1,\ldots,x_N) \, .
\end{equation}
Measurements of equal-time correlation functions represent, therefore, a powerful tool to quantify the combined effect of all possible quantum 
processes that contribute to an $N^{\mathrm{th}}$-order correlation --- no matter how high the order of a process in terms of powers of Planck's constant 
$h$ may be, or whether the contribution is of non-perturbative origin.\\

\subsection{Experimental correlation functions and their decompositions}
From the measured phase field $\varphi(z)$ we determine equal-time $N^\mathrm{th}$-order correlation functions
\begin{align}
G^{(N)}({\bf{z}},{\bf{z}}')\! =\! \braket{\triangle\varphi(z_1,z'_1)\dots\triangle\varphi(z_N,z'_N)} ~\mathrm{,}
\label{eq:CorrelationFunctionSI}
\end{align}
where $\triangle\varphi(z_i,z'_i)=\varphi(z_i)- \varphi(z'_i)$ are unambiguous phase differences of the unbound phase at different 
spatial points $z_i, z'_i$, and we suppressed the common time label $t$ to shorten the notation. 
In our experiment, the first-order correlation function vanishes by symmetry, $\braket{\triangle\varphi(z_i,z'_i)} = 0$, as well 
as all other correlation functions where $N$ is an odd positive integer. While all information is contained in the $N^\mathrm{th}$-order correlation functions, 
it is more enlightening to measure connected correlations, since the redundant information of lower-order correlations is eliminated as explained 
in the previous section. In the following we give explicit expressions for the decomposition 
\begin{align} \label{eq:factorizationGeneral}
 G^{(N)}({\bf{z}},{\bf{z}}') = G^{(N)}_{\mathrm{con}}({\bf{z}},{\bf{z}}') + G^{(N)}_{\mathrm{dis}}({\bf{z}},{\bf{z}}')
\end{align}
of the experimental correlations defined in \Eq{CorrelationFunctionSI}. 
The general formula for the connected part \cite{shiryaev2016probability} is
\begin{align}
G^{(N)}_{\mathrm{con}}({\bf{z}},{\bf{z}}') =  \sum_{\pi}{ \ (|\pi|-1)! \ (-1)^{|\pi|-1} \prod_{B \in \pi} \left\langle \prod_{i \in B} \triangle\varphi(z_i,z'_i) \right\rangle}. \label{SI_general_formula_connected} 
\end{align}
Here, the sum runs over all possible partitions $\pi$ of $\{1,\dots,n\}$, the first product runs over all blocks $B$ 
of the partition and the second product over all elements $i$ of the block. Since, in our system, all correlation functions where $N$ is an odd 
positive integer vanish by symmetry, we get $G_{\mathrm{con}}^{(2)}(\mathbf{z}, \mathbf{z}') = G^{(2)}(\mathbf{z}, \mathbf{z}')$ and, e.g., for the 
$4^\mathrm{th}$-order connected correlation function:
\begin{align}
\begin{split}
G_{\mathrm{con}}^{(4)}(\mathbf{z}, \mathbf{z}') = 
G^{(4)}(\mathbf{z}, \mathbf{z}')
&-
\braket{\triangle\varphi(z_1,z'_1)\triangle\varphi(z_2,z'_2)} 
\braket{\triangle\varphi(z_3,z'_3)\triangle\varphi(z_4,z'_4)} \\
&-
\braket{\triangle\varphi(z_1,z'_1)\triangle\varphi(z_3,z'_3)}
\braket{\triangle\varphi(z_2,z'_2)\triangle\varphi(z_4,z'_4)}
\\
&-
\braket{\triangle\varphi(z_1,z'_1)\triangle\varphi(z_4,z'_4)}
\braket{\triangle\varphi(z_2,z'_2)\triangle\varphi(z_3,z'_3)}
\,. 
\end{split}
\end{align}
In case of a Gaussian state, all connected parts of correlation functions $(N>2)$ vanish, i.e.\ 
$G^{(N>2)}_{\mathrm{con}}({\bf{z}},{\bf{z}}') \equiv 0$. Hence, all correlation functions factorise and one recovers Wick's 
theorem~\cite{zinn2002quantum} stating that, for a Gaussian state, all correlation functions with $(N>2)$ are determined by second-order correlation functions.
Explicitely, the Wick decomposition is given by 
\begin{align}
	G_{\mathrm{wick}}^{(N)}({\bf{z}},{\bf{z}}') =  \sum_{\pi_2} \Bigg[ \prod_{B \in \pi_2} \Big\langle &[\varphi(z_{B_1})- \varphi(z'_{B_1})]
	 [\varphi(z_{B_2})- \varphi(z'_{B_2})] \Big\rangle \Bigg] . 
	 \label{SI_eq:Wick_decomp}
\end{align}
Here the sum runs over all possible partitions $\pi_2$ of $\{1,\dots,n\}$ into blocks of size 2. The product again runs over all blocks $B$ of the partition (see \cite{shiryaev2016probability}).

The relevance of the connected part $G^{(N)}_{\mathrm{con}}$ can be quantified by the measure
\begin{align} \label{eq:measureM}
 M^{(N)} = \frac{\sum_{\bf{z}}{\left|G^{(N)}_{\mathrm{con}}({\bf{z}},0)\right|}}{\sum_{\bf{z}}{\left|G^{(N)}({\bf{z}},0)\right|}} 
\end{align}
with $M^{(N)} \in [0,1]$. For a Gaussian state, $M^{(N)} \equiv 0$ for all $N>2$.

Choosing coordinates $z_1 = z_2 = \ldots = z_N$ and $z'_1 = z'_2 = \ldots = z'_N$, the above formulas simplify and the $N^\mathrm{th}$-order 
connected correlation function can be determined by the recursion formula
\begin{align}
&G^{(N)}_{\mathrm{con}}({\bf{z}_1},{\bf{z}}_1') = G^{(N)}({\bf{z}}_1,{\bf{z}}'_1) - \sum_{m=1}^{N-1} 
\binom{N-1}{m-1} G^{(m)}_{\mathrm{con}}({\bf{z}}_1,{\bf{z}}_1') G^{(N-m)}({\bf{z}}_1,{\bf{z}}_1') \, .
\end{align}
Specifically for the lowest orders we get
\begin{align}
G^{(2)}_{\mathrm{con}}(\mathbf{z}_1,\mathbf{z}'_1) =& \braket{\triangle\varphi^2} \, , \notag \\
G^{(4)}_{\mathrm{con}}(\mathbf{z}_1,\mathbf{z}'_1) =& \braket{\triangle\varphi^4} - 3\braket{\triangle\varphi^2}^2 \, ,\notag \\ 
G^{(6)}_{\mathrm{con}}(\mathbf{z}_1,\mathbf{z}'_1) =& \braket{\triangle\varphi^6} - 15 \braket{\triangle\varphi^4} \braket{\triangle\varphi^2}^2 + 30 \braket{\triangle\varphi^2}^3  \, ,
\end{align}
with $\triangle\varphi = \triangle\varphi(z_1,z'_1)$. For a Gaussian state, we get from Wick's theorem
\begin{align}
G^{(N)}(\mathbf{z},\mathbf{z'_1}) &\stackrel{\mathrm{Gaussian}} = \, \langle \triangle\varphi^2 \rangle^N (N-1)!! \, ,
\end{align}
where $(\dots)!!$ is the double factorial. These simplified formulas will be helpful in the next section, where we discuss the factorisation 
properties of commonly used periodic observables.
\subsection{Connected versus periodic correlation functions}
The periodic observables used in our previous experiments \cite{Langen15},
\begin{align}\label{eq:PCF_GGE}
\mathcal{C}({\bf{z}},{\bf{z'}}) &\approx \braket{e^{i \sum_n \triangle\varphi(z_n,z'_n)}} \, ,
\end{align}
are, by neglecting the density fluctuations (suppressed due to atomic repulsion), related to correlations of the bosonic fields 
$\psi_{1,2}$ via
\begin{align}\label{eq:corr_GGE}
 \mathcal{C}({\bf{z}},{\bf{z'}}) :&= \frac{\langle \psi_1(z_1)\psi^{\dagger}_2(z_1)\psi^{\dagger}_1(z'_1)\psi_2(z'_1) \dots \psi_1(z_N)\psi^{\dagger}_2(z_N)\psi^{\dagger}_1(z'_N)\psi_2(z'_N) \rangle}{\sum_{n=1}^N \sqrt{\langle |\psi_1(z_n)|^2 \rangle \langle |\psi_2(z_n)|^2 \rangle \langle |\psi_1(z'_n)|^2 \rangle \langle |\psi_2(z'_n)|^2 \rangle}} \, .
\end{align}
These correlations are not suitable for the present analysis as even the second-order correlation function 
$\mathcal{C}(z_1,z'_1) = \braket{e^{i \triangle\varphi(z_1,z'_1)}}$ contains all higher cumulants of the 
phase. This can be directly seen by expanding the expression
$\log \braket{e^{i\lambda \triangle\varphi(z_1,z'_1)}}$
(usually called the cumulant generating function) in powers of $\lambda$ resulting in
\begin{align}
\log \braket{e^{i\lambda \triangle\varphi}} &=  i \lambda    \braket{\triangle\varphi} 
- \frac{\lambda^2 }{2} \left\{ \braket{\triangle\varphi^2} -
 \braket{\Delta \varphi}^2 \right\}
+ \frac{(i\lambda)^3 }{3!} \left\{  \braket{\triangle\varphi^3} 
- 3  \braket{\triangle\varphi} \braket{\triangle\varphi^2} +  2 \braket{\triangle\varphi}^3 \right\} \notag \\
&- \frac{\lambda^4 }{4!} \left\{ \braket{\triangle\varphi}^4
- 4 \braket{\triangle\varphi} \braket{\triangle\varphi^3}
-3 \braket{\triangle\varphi^2}^2 + 12 \braket{\triangle\varphi^2}
\braket{\triangle\varphi}^2 -6  \braket{\triangle\varphi}^4\right\} + \mathcal{O}(\lambda^5)\, \notag \\
&= \exp \left[ \sum_{m=1}^{\infty} \frac{(-\lambda)^m}{(2m)!}{G^{(2m)}_{\mathrm{con}}(z_1,z'_1)} \right] ~\mathrm{,}
\end{align} 
where, again, $\triangle\varphi = \triangle\varphi(z_1,z'_1)$. For $\lambda = 1$ one recovers the second-order correlation function 
\begin{align}\label{eq:GGE2pointcorrexpanded}
\mathcal{C}(z_1,z'_1)  = \exp \left[ \sum_{m=1}^{\infty} 
\frac{(-1)^m}{(2m)!}{G^{(2m)}_{\mathrm{con}}(z_1,z'_1)} \right] \, .
\end{align}
This shows that it is experimentally not possible to extract the individual connected correlation functions 
$G^{(N)}_{\mathrm{con}}$, i.e.\ information about $N$-particle interactions (see the following section), from the periodic correlations \eqref{eq:PCF_GGE}.
In case of Gaussian fluctuations of the phase, \Eq{GGE2pointcorrexpanded} reduces to
\begin{align}\label{eq:exp_2point}
\mathcal{C}(z_1,z'_1) = \exp \left[ -\frac{1}{2} \braket{[\varphi(z_1) - \varphi(z'_1)]}^2 \right] \, ,
\end{align}
with the correct factorisation, e.g.\ for the $4^\mathrm{th}$-order correlation function, given by
\begin{align}
 \mathcal{C}(z_1,z_2,z'_1,z'_2) = \frac{\mathcal{C}(z_1,z'_1) \mathcal{C}(z_2,z'_2) \mathcal{C}(z'_1,z_2) \mathcal{C}(z'_2,z_1))}{\mathcal{C}(z_1,z_2) \mathcal{C}(z'_1,z'_2)} \, .
\end{align}
This form of the factorisation, determined from Gaussian fluctuations of the phase, is due to the periodicity and the resulting restricted (finite) domain of these correlation functions.
\subsection{Relation to quasiparticle interactions}

In this section, we give a brief explanation as to how higher-order connected correlation functions are related to quasiparticle interactions, 
i.e.\ fully connected diagrams. Be aware that we do not anticipate to solve the problem using perturbation theory in any way, and hence 
do not concern ourselves with the inevitable problems of divergencies occurring in the perturbative expansion, and their solutions using 
well-established field-theoretical tools as resummation, renormalisation, and summation of divergent series. For details of the presented methods see any 
book on (statistical) field theory, e.g.\ \cite{zinn2002quantum}.

In thermal equilibrium the equal-time correlation functions defined in \Eq{CorrelationFunctionSI} are, due to the linearity of the trace, determined by correlations 
of the form
\begin{equation}\label{eq:corrQP}
 \Big\langle \prod_{i=1}^{N} \varphi(z_i) \Big\rangle \equiv Z^{-1}{\operatorname{Tr} \big[ \e^{-\beta H} \prod_{i=1}^{N} \varphi(z_i) \big]} ~\mathrm{,}
\end{equation}
where $Z=\operatorname{Tr} [ \e^{-\beta H} ]$ is the partition function, and the trace is defined as $\operatorname{Tr}[\dots] = \sum_n \langle n|\dots|n\rangle$, 
with $|n\rangle$ being a complete, orthonormal basis of the Hilbert space. While being exact this equation is in general not solvable without further 
approximations. First, we will approximate the system through its low-energy effective theory, discussed in Sect.~\ref{sec:sGlEeffTh}, and therefore 
determined by the sine-Gordon Hamiltonian \eqref{eq:SGextended}. Expanding the cosine we write $H_{\mathrm{SG}}$ as
\begin{align}
 H &= \int dz : \left[ g \delta {\rho}^2 + 
\frac{\hbar^2 n_\mathrm{1D}}{4m} \left( \frac{\partial {\varphi}}{\partial z}\right)^2  + \hbar \tilde{J} n_\mathrm{1D} \varphi^2 \right] :  
-  \int dz : \left[ 2\hbar \tilde{J} n_{\mathrm{1D}} \sum_{n=2}^{\infty} \frac{(-1)^n}{(2n)!} \varphi^{2n} \right] : \\
   &= H_0 + V ~\mathrm{,}
\end{align}
where we split the Hamiltonian into a free part $H_0$, quadratic in the fields, and an interaction part $V = V_4 + V_6 + \dots$, containing all higher-order terms. 
We wrote the Hamiltonian in its normal-ordered form (where all creation operators are to the left, denoted by $:\,:$) which leads to a multiplicative renormalisation of the 
coupling $J \to \tilde{J}$, and we dropped an irrelevant constant (see e.g.\ Ref.~\cite{Coleman75}).

Diagonalisation of the quadratic Hamiltonian $H_0$ defines the quasi-particle basis through the 
Bogoliubov expansion
\begin{align}
 B(z) = \sum_k \big[ u_k(z) \, b_k + v_k^*(z) \, b_k^{\dagger} \big]
\end{align}
of the quadrature field $B(z) = \varphi(z) \sqrt{n_{\mathrm{1D}} / 2} - \i \delta\rho(z) / \sqrt{2 n_{\mathrm{1D}}}$. The mode functions 
$(u_k, \, v_k)$ are eigenmodes of the Bogoliubov operator with eigenvalues $\epsilon_k$. They ensure cancellation of all non-diagonal quadratic terms 
in the Hamiltonian and are normalized as $\int \, \mathrm{d}z[ |u_k(z)|^2 - |v_k(z)|^2] = 1$. Finally, the quadratic Hamiltonian 
takes the form
\begin{align}
 H_0 = \sum_k \epsilon_k b_k^{\dagger} b_k ~\mathrm{,}
\end{align}
describing non-interacting quasiparticles. The mode expansion of the fields is given by
\begin{align}
 \delta\rho(z) &= \sum_k \delta\rho_k(z) \, b_k + \delta\rho_k^*(z) \, b_k^{\dagger} \\
 \varphi(z) &= \sum_k \varphi_k(z) \, b_k + \varphi_k^*(z) \, b_k^{\dagger} ~\mathrm{,}
\end{align}
where $\delta\rho_k(z) = [u_k(z) + v_k(z)] \sqrt{n_{\mathrm{1D}}}$ and $\varphi_k(z) = [u_k(z) - v_k(z)]/(2 \i \sqrt{n_{\mathrm{1D}}})$.
In the limiting cases of zero or very strong tunnel coupling the interaction potential $V$ vanishes, for the latter due to smallness of the fluctuations $\varphi$. 
Inserting the mode expansion into \Eq{corrQP}, 
a direct calculation of the trace in the quasiparticle Fock basis leads to the observed factorisation of correlations, as expected for a free theory.
This can easily be generalised to arbitrary order by use of  Wick's theorem for thermal states. Factorisation of equal-time phase correlation functions according to Wick's theorem, as was determined in the experiment for the uncoupled and the
strongly coupled system, therefore  shows the absence of quasiparticle interactions in the theory. Note that this is by far not a trivial result, even for vanishing coupling $J$, 
as we neglected an infinite number of higher order terms by replacing the full Hamiltonian $H$ by the low-energy effective model 
$H_{\mathrm{SG}}$.

In case of a non-vanishing interaction potential $V$ the equations become increasingly more complicated due to the non-vanishing commutator $[H_0,V]$.
In thermal equilibrium, the correlation functions of the phase can be calculated in perturbation theory in the imaginary-time (Matsubara) formalism.
One defines the time-ordered correlation functions in imaginary time $\tau$ (Matsubara Green's functions)
\begin{align}\label{eq:MatsubaraGreen}
 \langle \hat{T} \varphi_H(\tau_1,z_1) \dots \varphi_H(\tau_N,z_N) \rangle \equiv \frac{\operatorname{Tr} \big[ \e^{-\beta H_0} \, \hat{T} \, \mathcal{U}(\beta,0) \varphi_I(\tau_1,z_1) \dots \varphi_I(\tau_N,z_N) \big]}{\operatorname{Tr} \big[ \e^{-\beta H_0} \, \mathcal{U}(\beta,0) \big]} ~\mathrm{,}
\end{align}
where $\varphi_H(\tau,z) = \e^{\tau H} \varphi(z) \e^{-\tau H} $ are the Heisenberg field operators in imaginary time $\tau$, and $\varphi_I(\tau,z) = \e^{\tau H_0} \varphi(z) \e^{-\tau H_0}$ are the fields in the interaction picture (denoted by the subscript $I$), evolving in imaginary time with the free Hamiltonian $H_0$.
The time evolution operator $\mathcal{U}(\beta,0)$ fulfills
\begin{align}
 \partial_{\tau} \mathcal{U}(\tau,0) = - V_I(\tau) \mathcal{U}(\tau,0) ~\mathrm{.}
\end{align}
It can be written as the Dyson series 
\begin{align}
 \mathcal{U}(\tau,\tau') = \hat{T} \e^{-\int_{\tau'}^{\tau} \mathrm{d}\tau'' V_I(\tau'') } = \sum_{n=0}^{\infty} \frac{(-1)^n}{n!} \int_{\tau'}^{\tau} \mathrm{d}\tau_1 \dots \int_{\tau'}^{\tau} \mathrm{d}\tau_n \, \hat{T}\, V_I(\tau_1) \dots V_I(\tau_n) ~\mathrm{,}
\end{align}
which allows to express the correlation functions \eqref{eq:MatsubaraGreen}, up to any order in $V_I$, as a diagrammatic expansion in Feynman diagrams.
The sine-Gordon Hamiltonian, \Eq{SGextended}, represents a scalar field theory with an infinite number of polynomial interaction terms.
Standard results of quantum field theory allow to distinguish between three distinct types of diagrams. First, all diagrams in which the fields of the interaction potential $V$
are contracted among themselves and are otherwise disconnected (vacuum diagrams). These vacuum diagrams are exactly canceled by the denominator in \Eq{MatsubaraGreen} to all orders in the perturbative expansion. Second, all diagrams which are not fully connected only 
contribute to the disconnected part of the correlation function, and can be factorised into full, lower-order correlation functions. Third, the fully connected diagrams 
describe genuine $N$-body quasiparticle interactions and constitute the connected part of the correlation function.

\footnotetext[3]{Note that, in general, the limit $\tau_1,\dots,\tau_N \to \tau$ need to be taken with care as the time-ordered Matsubara Greens-functions due to non-vanishing commutators might be discontinuous at equal times (see also footnote 2). Since, for the correlations considered here, the equal-time commutator vanishes no further problems occur in taking the equal-time limit.}

Note that the above time-ordered imaginary-time correlation functions are only related to physical observables for equal times $\tau_1 = \dots = \tau_N$,
for which they coincide with the experimentally measured correlation functions.\footnotemark[3] However, the Matsubara Green's functions may be 
analytically continued to the real-time axis to determine the physically relevant retarded Green's functions. This continuation immediately allows 
to infer the effect of $N$-particle interactions in the theory. As explained in \Sect{equal-time-corr}, the $N^{\mathrm{th}}$-order 
equal-time connected correlation function represents the sum over all these fully connected processes with $N$ external lines. 
Since the experimentally measured phase fields are linear in the quasiparticle creation/annihilation operators, $N^{\mathrm{th}}$-order correlation functions 
are a direct measure for the combined effect of the $N$-body quasiparticle interaction (to all orders in the coupling). This is in contrast 
to the periodic observables, discussed in the previous section, which sum over all possible quasiparticle interactions (for all values of $N$).
Measurements of higher-order correlation functions therefore allow for a direct comparison to highly non-trivial field-theoretical calculations, and give valuable 
information about the convergence of the perturbative expansion, the validity of non-perturbative theoretical methods, and the summation of divergent series.

%
%
%

\end{document}